\documentclass[twocolumn,showkeys]{aastex62}

\usepackage{longtable}
\usepackage{graphicx}
\usepackage{amsmath,amssymb}
\usepackage{color}
\usepackage{units}
\usepackage{epstopdf}
\usepackage{hyperref}
\usepackage{multirow}
\usepackage{url}
\usepackage{placeins}

\usepackage{rotating}

\usepackage{subfigure}

\newcommand{\beq}{\begin{equation}}
\newcommand{\eeq}{\end{equation}}
\newcommand{\bdm}{\begin{displaymath}}
\newcommand{\edm}{\end{displaymath}}

\definecolor{Gray}{gray}{0.9}
\definecolor{orange}{rgb}{0.9,0.5,0}

\newcommand{\beqn}{\begin{eqnarray}}
\newcommand{\eeqn}{\end{eqnarray}}

\graphicspath{{./plots/}}

\begin{document}

\title{Implications of the search for optical counterparts during the first six months of the Advanced LIGO's and Advanced Virgo's third observing run: possible limits on the ejecta mass and binary properties}

\author[0000-0002-8262-2924]{Michael W. Coughlin}
\affiliation{California Institute of Technology, 1200 East California Blvd, MC 249-17, Pasadena, CA 91125, USA}
\author[0000-0003-2374-307X]{Tim Dietrich}
\affiliation{Nikhef, Science Park, 1098 XG Amsterdam, The Netherlands}
\author[0000-0002-7686-3334]{Sarah Antier}
\affiliation{APC, UMR 7164, 10 rue Alice Domon et Léonie Duquet, 75205 Paris, France}
\author{Mattia Bulla}
\affiliation{Nordita, KTH Royal Institute of Technology and Stockholm University, Roslagstullsbacken 23, SE-106 91 Stockholm, Sweden}
\affiliation{Oskar Klein Centre, Department of Physics, Stockholm University, SE-106 91 Stockholm, Sweden}
\author{Francois Foucart}
\affiliation{Department of Physics \& Astronomy, University of New Hampshire, 9 Library Way, Durham NH 03824, USA}
\author{Kenta Hotokezaka}
\affiliation{Department of Astrophysical Sciences, Princeton University, Princeton, NJ 08544, USA}
\author{Geert Raaijmakers}
\affiliation{GRAPPA, Anton Pannekoek Institute for Astronomy and Institute of High-Energy Physics,
University of Amsterdam, Science Park 904, 1098 XH Amsterdam, The Netherlands}
\author{Tanja Hinderer}
\affiliation{GRAPPA, Anton Pannekoek Institute for Astronomy and Institute of High-Energy Physics,
University of Amsterdam, Science Park 904, 1098 XH Amsterdam, The Netherlands}
\author{Samaya Nissanke}
\affiliation{GRAPPA, Anton Pannekoek Institute for Astronomy and Institute of High-Energy Physics,
University of Amsterdam, Science Park 904, 1098 XH Amsterdam, The Netherlands}
\affiliation{Nikhef, Science Park 105, 1098 XG Amsterdam, The Netherlands}

\begin{abstract}
GW170817 showed that neutron star mergers not only emit gravitational waves but also can release electromagnetic signatures in multiple wavelengths. 
Within the first half of the third observing run of the Advanced LIGO and Virgo detectors, there have been a number of gravitational wave candidates of compact binary systems for which at least one component is potentially a neutron star. 
In this article, we look at the candidates S190425z, S190426c, S190510g, S190901ap, and S190910h, predicted to have potentially a non-zero remnant mass, in more detail. All these triggers have been followed up with extensive campaigns by the astronomical community doing electromagnetic searches for their optical counterparts; however, according to the released classification, there is a high probability that some of these events might not be of extraterrestrial origin. 
Assuming that the triggers are caused by a compact binary coalescence and that the individual source locations have been covered during the EM follow-up campaigns, we employ three different kilonova models and apply them to derive possible constraints on the matter ejection consistent with the publicly available gravitational-wave trigger information and the lack of a kilonova detection. These upper bounds on the ejecta mass can be related to limits on the maximum mass of the binary neutron star candidate S190425z and to constraints on the mass-ratio, spin, and NS compactness for the potential black hole-neutron star candidate S190426c. Our results show that deeper electromagnetic observations for future gravitational wave events near the horizon limit of the advanced detectors are essential.
\end{abstract}



\section{Introduction}

By the combined detection of GW170817, AT2017gfo, and GRB170817A, 
the field of multi-messenger astronomy was ushered into a new era in which gravitational-wave (GW) and electromagnetic (EM) signatures are simultaneously measured and analyzed, e.g.,~\cite{AbEA2017b,GBM:2017lvd,ArHo2017,2017Sci...358.1556C,LiGo2017,MoNa2017,SaFe2017,SoHo2017,TaLe2013,TrPi2017,VaSa2017}.
Joint analyses allow a better understanding of the supranuclear-dense matter inside neutron stars (NSs) (e.g.~\cite{RaPe2018,RaDa2018,BaJu2017,MaMe2017,ReMo2017,CoDi2018,CoDi2018b,Capano:2019eae}), a precise measurement of the speed of gravitational waves~\citep{AbEA2017e}, an independent measurement of the expansion rate of the Universe~\citep{AbEA2017g,Hotokezaka:2018dfi,Coughlin:2019vtv,DhBu2019}, and constraints on alternative models of gravity \citep{EzMa17,BaBe17,CrVe17,SaJa17}. 

In general, the merger of two compact objects from which at least one is a NS, is connected to a variety of possible EM signatures in almost all wavelengths. A highly relativistic jet can produce a short gamma-ray burst (sGRB) lasting a few seconds \citep{1989Natur.340..126E, Paczynski1991, 1992ApJ...395L..83N, MocHer1993, LeeRR2007, Nakar2007} and a synchrotron afterglow in the X-rays, optical and radio visible bands for hours to months after the initial emission due to the deceleration of the jet into the ambient media \citep{SaPiNa1998}. 
The ejection of highly neutron rich material, being the seed of r-process elements \citep{LaSc1974, LaSc1976}, powers a thermal ultraviolet/optical/near-infrared kilonova due to the radioactive decay of the new heavy elements produced in the ejecta \citep{LiPa1998,MeMa2010,RoKa2011,KaMe2017}. Although the color and luminosity of a kilonova will be viewing angle dependent, the kilonova signature is, in contrast to the sGRB and its afterglow, likely visible from all viewing angles. 
This means that after every merger which ejects a sufficient amount of material, one should be able to observe a kilonova regardless of the orientation of the system \citep{RoKa2011}. Thus, kilonovae provide a \textit{smoking guns} evidence for binary neutron star (BNS) and black hole - neutron star (BHNS) mergers.\\

However, current numerical relativity studies indicate that not all BNS or BHNS collisions will eject enough material to create EM signals as bright as the one observed for GW170817. For most BNS systems, the EM signals are expected to be dimmer than for GW170817 if a black hole (BH) forms directly after the moment of merger, since for these \textit{prompt collapse} configurations the amount of ejected material and the mass of the potential debris disk is expected to be very small. Whether a merger remnant undergoes a prompt collapse depends mostly on its total mass~\citep{Bauswein:2013jpa,HoKi13,DiUj2017,Koppel:2019pys,Agathos:2019sah} but also seems to be sub-dominantly affected by the mass-ratio~\citep{Kiuchi:2019lls}. For highly asymmetric mass ratios ($m_1/m_2\lesssim 0.8$), there could be a non-negligible ejecta mass and/or a massive accretion disk around the black hole remnant even for prompt collapse scenarios~\citep{Kiuchi:2019lls}.

In the case of a BHNS system, the brightness of the potential EM counterpart depends on whether the NS gets tidally disrupted by the BH and, thus, ejects a large amount of material and forms a massive accretion disk; or if the star falls into the BH without disruption, preventing the production of GRBs and kilonovae. 
Thus, the outcome of the merger is mostly determined by the mass ratio of the binary, the spin of the black hole, and the compactness of the NS, with disruption being favored for low-mass, rapidly rotating BH and large NS radii~\citep{Etienne:2008re,Pannarale:2010vs,Foucart:2012nc,Kyutoku:2015gda,Kawaguchi:2016ana,Foucart:2018rjc}.

Since the beginning of the third observation run, a number of potential GW events have triggered extensive follow-up campaigns to search for possible EM counterparts, most notably S190425z~\citep{SiEA2019a,SiEA2019b}, S190426c~\citep{ChEA2019a,ChEA2019b}, S190510g~\citep{gcn24442}, S190814bv~\citep{gcn25324}, S190901ap~\citep{gcn25606}, S190910h~\citep{gcn25707}, S190910d~\citep{gcn25695}, S190923y~\citep{gcn25814}, and S190930t~\citep{gcn25876}; cf.\ Tab.~\ref{tab:allevents} for more details.\footnote{Additional alerts have been sent out for other triggers, but those have been retracted. A BNS candidate S190718y \citep{gcn25087} was sent to the astronomical community; due to the presence of a strong glitch near to the trigger time, only a few optical observations were performed and this alert will not be considered in this study. In addition, other candidates S190518bb \citep{gcn24591}, S190524q \citep{gcn24656}, S190808ae \citep{gcn25296}, S190816i \citep{gcn25367} and S190822c \citep{gcn25442} were also identified and later retracted. In addition, an interesting black hole merger candidate triggered intensive follow-up due to its low latency properties results with the possibility to have one object between 3 and 5 solar mass \citep{gcn25187}, but updated results with the full exploration of the parameter space of masses and spins, finally did not confirm these properties \citep{gcn25208}.} 
The large size of localization regions with thousands of square degrees have proved much more challenging to cover over short times than the $\sim 20$ square degrees of GW170817. In fact, no joint detection of GW and EM signals have been confirmed; see also \cite{DaDa:2019} for a possible explanation that no sGRBs has been observed for the GW events within O3a.
While a detection of an EM signature will help significantly to unravel some of the remaining open questions related to compact binary mergers, the possibility of a ``missing'' EM signature for an astrophysical relevant trigger whose sky location was covered during an EM follow-up campaign also delivers some information about the source properties, as we will discuss. \\

\begin{table}
  \centering
  \caption{Overview about officially non-retracted GW triggers
  with large probabilities to be BNS or BHNS systems. 
  The individual columns refer to: 
  the name of the event, an estimate using the most up-to-date classification for the event to be 
  a BNS [p(BNS)], a BHNS [p(BHNS)], or terrestrial noise [p(terrestrial)] \citep{KaCa2019}, 
  and an indicator to estimate the probability of producing EM signature considering the candidate with astrophysical origin [p(HasRemnant)], whose definition is in the \href{https://emfollow.docs.ligo.org/userguide/content.html}{LIGO-Virgo alert userguide}. Note that the alert can be also classified as ``MassGap,'' completing the possible classifications. 
  Note that within our analysis, we do not consider S190718y
  because of its very low probability to be of astrophysical origin. }
\begin{tabular}{l|cccc}
\hline
Name      & p(BNS)   & p(BHNS) & p(terr.) & p(HasRemn.) \\
\hline
S190425z  & $> 99\%$ & $0\%$    &   $< 1\%$      & $> 99\%$    \\  
S190426c  & $24\%$   & $6\%$    &    $58\%$       & $> 99\%$   \\ 
S190510g  & $42\%$   & $0\%$    &    $58\%$      & $> 99\%$    \\ 
S190718y$^*$& $2\%$    & $0\%$    &    $98\%$    & $> 99 \%$     \\
S190814bv & $0\%$    & $> 99\%$ &   $< 1\%$    & $< 1\%$       \\ 
S190901ap & $86\%$   & $0\%$    &    $14\%$    & $> 99\%$      \\ 
S190910d  & $0\%$    & $98\%$   &      $2\%$     & $< 1\%$     \\ 
S190910h  & $61\%$   & $0\%$    &      $39\%$    & $> 99\%$    \\ 
S190923y  & $0\%$    & $68\%$   &      $32\%$    & $< 1\%$     \\
S190930t  & $0\%$    & $74\%$   &      $26\%$    & $< 1\%$         
\end{tabular}
 \label{tab:allevents}
\end{table}

In this article, we try to understand if from the detection or, more likely, non-detection of an EM counterpart to a potential GW event it is possible to place constraints on the merger outcome and the properties of the system. For this purpose, we will shortly summarize the EM follow-up campaigns of S190425z, S190426c, S190510g, S190814bv, S190901ap, S190910d, S190910h, S190923y, and S190930t in Sec.~\ref{sec:EM_follow_up_campaigns}. 
We further also refer to~\cite{AnGo:2019} for a dedicated discussion done by the GROWTH collaboration about S190814bv. 

In Sec.~\ref{sec:limits} we focus on the events for which the \texttt{HasRemnant}\footnote{https://emfollow.docs.ligo.org/userguide/content.html and https://dcc.ligo.org/LIGO-P1900291; Typically, the \texttt{HasRemnant} classification employs the disk mass estimate of~\cite{Foucart:2018rjc} and applies to BHNS systems. BNS configurations are assumed to cause an EM signature, which, as we show later, might not be correct. The \texttt{HasRemnant} classification assumes the event to be of astrophysical origin and does not incorporate the possibility that the trigger is caused by noise} prediction provides a high probability of a potential EM signature (S190425z, S190426c, S190510g, S190901ap, and S190910h)\footnote{We do not include S190718y because of its high probability to be noise.}; cf.~Tab.~\ref{tab:allevents}. Under the assumption that the GW candidate location was covered during the EM observations, we will use a set of three different lightcurve models~\citep{KaMe2017,Bul2019,HoNa2019} to predict the properties of the kilonova consistent with the non-observation of an EM counterpart. This analysis allows us to derive constraints on the maximum ejecta mass for each event in Sec.~\ref{sec:limits} and connects our findings to the binary properties in Sec.~\ref{sec:binary_property}. 
These constraints are typically not very striking given the large distance to the GW triggers in the first half of advanced LIGO and advanced Virgo's third observing run, which highlights that, if possible, longer exposure times should be employed to reduce the possibility that interesting transients might be missed. We summarize our conclusions and lessons learned for observations in the second half of the third observing run in Sec.~\ref{sec:summary}.

\section{EM follow-up campaigns}
\label{sec:EM_follow_up_campaigns}

We summarize the EM follow-up work of the various teams that performed synoptic coverage of the sky localization area and who have circulated their findings in publicly available circulars during the first six months of the third observing run.
For a summary of the follow-up campaign during the second observing run, please see \cite{AbEA2019} and references therein.
We differentiate the candidates by their classification (predominantly BNS in Table~\ref{tab:TableobsBNS} and predominantly BHNS in Table~\ref{tab:TableobsNSBH}).
While this is mostly an initial classification and may change based on future offline estimates, we think it is useful as, for example, the distance estimates tend to be different between these classes.
A short discussion about each candidate is presented below; note that we do not report the observations that exclusively target galaxies.

\subsection{\textbf{S190425z}}

LIGO/Virgo S190425z was identified by the LIGO Livingston Observatory (L1) and the 
Virgo Observatory (V1) at 2019-04-25 08:18:05.017 UTC \citep{SiEA2019a,SiEA2019b}. LIGO Hanford Observatory (H1) was not taking data at the time. It has been so far categorized as a BNS signal, reported as a BNS $(99\%)$ with a small probability of being in the mass gap $(<1\%)$. Due to the low signal-to-noise ratio (SNR) in V1, S190425z's sky localization is relatively poor, covering nearly $10,000$ square degrees.
The original distance quoted for this system is $155 \pm 45$\,Mpc, thus, about $\sim 4$ times further away than GW170817.

As the first alert during the O3 campaign with a high probability of having a counterpart, there was an intense follow-up campaign within the first $\sim$\,72 hours after the initial notice (see $\approx$ 120 reports in \href{https://gcn.gsfc.nasa.gov/other/GW190425z.gcn3}{GCN archive}, mostly focusing on optical follow-up). 
As expressed in \citet{gcn24232}, with more than 50,000 galaxies compatible with the 90\% sky area volume due to the large uncertainty of the localization, it was difficult to fully cover S190425z's localization. However, as shown in Table~\ref{tab:TableobsBNS}, ten telescopes reported tiling observations of the localization. For example, both the Zwicky Transient Facility (ZTF) \citep{Bellm2018,Graham2018,DeSm2018,MaLa2018}, a camera and associated observing system on the Palomar 48 inch telescope, and Palomar Gattini-IR, a new wide-field near-infrared survey telescope at Palomar observatory, followed up S190425c extensively \citep{CoAh2019b}. Covering about 8000 and 2200 square degrees respectively, the systems achieved depths of $\approx$\,21 $m_\textrm{AB}$ in g- and r-bands with ZTF and 15.5\,mag in J-band with Gattini-IR. Among them, using the LALInference skymap, about 21\% of and 19\% of the sky localization was covered by ZTF and Palomar Gattini-IR respectively. In addition, Pan-STARRS covered 28\% of the bayestar sky localization area in g-band with a limiting magnitude of $i=21.5$\,mag \citep{gcn24210}; similarly, GOTO covered 30\% of the initial skymap down to $L=20.5$\,mag \citep{gcn24224}.

\subsection{\textbf{S190426c}}

LIGO/Virgo S190426c was identified by H1, L1, and V1 at 2019-04-26 15:21:55.337 UTC \citep{ChEA2019a,ChEA2019b}. 
With a probability of 58\% to be terrestrial, S190426c might not be of astrophysical origin. 
But assuming that the signal is of astrophysical relevance, S190426c seems to be a BHNS system with relative probabilities of approximately 12 : 5 : 3 : 0 for the categories NSBH : MassGap : BNS : BBH, respectively~\citep{gcn24411}. Within this analysis the \texttt{HasRemnant} probability is stated as $72\%$, thus, for all events with large \texttt{HasRemnant} predictions, is our best example for a possible BHNS merger. 
S190426c's sky localization, given that it was discovered by multiple interferometers, covers less area than S190425z. The initial 90\% credible region was 1260 deg$^2$ with a luminosity distance of $375 \pm 108$\,Mpc \citep{ChEA2019a}.
The updated skymap, sent 48\,hrs after the initial skymap, had a 90\% credible region of 1130 deg$^2$ and a luminosity distance estimate of $377 \pm 100$\,Mpc \citep{gcn24279}. As the first event announced with a significant probability of a BHNS nature, the interest in this event was large and about 70 circulars have been sent out (see the \href{https://gcn.gsfc.nasa.gov/other/GW190426c.gcn3}{GCN archive}). As shown in Table~\ref{tab:TableobsNSBH}, 13 telescopes scanned the localization region; for example, ASAS-SN \citep{gcn24309}, GOTO \citep{gcn24291}, and ZTF \citep{gcn24283} covered more than 50\% of the sky localization area using multiple filters in the first 48\,hrs.

\subsection{\textbf{S190510g}}
LIGO/Virgo S190510g was identified by H1, L1 and V1 at 2019-05-10 02:59:39.292 UTC \citep{gcn24442}. S190510g's latest sky localization covers 1166 deg$^2$ with a luminosity distance of $227 \pm 92$\,Mpc \citep{gcn24448}. In the most recent update provided by the LIGO and Virgo Collaboration, the event is now more likely caused by noise \citep{gcn24489} than it is to be an astrophysical source, with a probability of terrestrial (58\%) and BNS (42\%); however, since the event is, up to now, not officially retracted, we will consider it in this article. Due to its potential BNS nature and its trigger time being close to the beginning of the night in the Americas, the event was followed-up rapidly, with about 60 circulars produced (see \href{https://gcn.gsfc.nasa.gov/other/GW190510g.gcn3}{GCN archive}). With $\sim$\,65\% coverage of the LALInference skymap, GROWTH-DECam realized the deepest follow-up \citep{AnGo2019}. 
We can compute the joint coverage of different telescopes based upon their pointings and field of view reporting in the GCNs. Within 24\,hr, CNEST, HMT, MASTER, Xinglong and TAROT, all with clear filters down to 18\,mag, observed 71\% of the LALInference sky localization area; this number would assuredly be higher with a coordinated effort.

\subsection{\textbf{S190814bv}}
The candidate S190814bv was identified by H1, L1, and V1 on 2019-08-14 21:10:39.013 UTC. First classified as a compact merger with one component having an initial mass between 3 and 5 solar masses \citep{gcn25324}, the candidate is now classified as a BHNS with posterior support from parameter estimation \citep{VeRa2015} with NSBH ($>$99\%) \citep{gcn25333}. Initially, two different Bayestar-based sky localizations were generated, one with the lower false alarm rate which included Livingston and Virgo data (sent 21\,min after the trigger time) and one with contribution of the three instruments (sent 2\,hr after the GW trigger time). A third skymap (LALInference) with all three interferometers was sent $\sim$\,13.5\,hr after the trigger time. The initial three interferometer 90\% credible region was 38 deg$^2$ with a luminosity distance estimated at $276 \pm 56$\,Mpc. The latest 90\% credible region is 23 deg$^2$ with a luminosity distance of $267 \pm 52$\,Mpc. With the small localization region, and its location in the Southern hemisphere, the event was ideal for follow-up. However, no counterpart candidates remain after the extensive follow-up, with about 70 circulars produced (see \href{https://gcn.gsfc.nasa.gov/other/GW190814bv.gcn3}{GCN archive}).
As shown in Table~\ref{tab:TableobsNSBH}, many survey systems covered a vast majority of the localization region, including ATLAS \citep{gcn25375}, DESGW-DECam \citep{gcn25336}, and TAROT \citep{gcn25338}.
We note here despite the small sky area and the intensive followed-up studies, we do not consider this object in the analysis due to its \texttt{HasRemnant} value.The joint coverage of MASTER and TAROT with 17 mag in clear filter within the first 3 hours was about 90\% of the LALinference skymap.

\subsection{\textbf{S190901ap}}
LIGO/Virgo S190901ap was identified by L1 and V1 at 2019-09-01 23:31:01.838 UTC \citep{gcn25606}. The candidate is currently classified as BNS (86\%) and terrestrial (14\%). The latest 90\% credible region is 14753 deg$^2$ with a luminosity distance of $241 \pm 79$\,Mpc \citep{gcn25614}, whereas the initial 90\% credible region was 13613 deg$^2$ with a luminosity distance of $242 \pm 81$\,Mpc.
Although considered as an interesting event due to a possible remnant, the large error box of thousands of square degrees led to a bit less interest in following-up the event (see $\approx$ 44 reports in \href{https://gcn.gsfc.nasa.gov/other/GW190901ap.gcn3}{GCN archive}). However, survey instruments such as GOTO \citep{gcn25654}, ZTF \citep{gcn25616} and MASTER \citep{gcn25609} observed more than 30\% of the localiztion; in particular, ZTF covered more than 70\%.

\subsection{\textbf{S190910d}}

LIGO/Virgo S190910d was identified as a compact binary merger candidate by H1 and L1 at 2019-09-10 01:26:19.243 UTC \citep{gcn25695}.
The candidate is currently classified as NSBH (98\%) and terrestrial (2\%). With an initial 90\% credible region of 3829
deg$^2$ with a luminosity distance of $606 \pm 197$\,Mpc, the latest 90\% credible region is 2482 deg$^2$ with a luminosity distance of $632 \pm 186$\,Mpc \citep{gcn25723}.
Relatively few instruments participated in the follow-up of this object (see $\approx$ 25 reports in \href{https://gcn.gsfc.nasa.gov/other/GW190910d.gcn3}{GCN archive}).
However, network instruments such as ZTF \citep{gcn25706}, GRANDMA-TAROT \citep{gcn25749}, and MASTER \citep{gcn25694} observed 25\% of the skymap or more. 

\subsection{\textbf{S190910h}}

LIGO/Virgo S190910h was identified as a compact binary merger candidate by only one detector (L1) at 2019-09-10 08:29:58.544 UTC \citep{gcn25707}.
The candidate is currently classified as BNS (61\%) and terrestrial (39\%). The initial 90\% credible region was 24226 deg$^2$ with a luminosity distance of $241 \pm 89$\,Mpc.
The latest 90\% credible region is 24264 deg$^2$ with a luminosity distance of $230 \pm 88$\,Mpc \citep{gcn25778}. Even fewer instruments participated in the follow-up of this object (see $\approx$ 20 reports in \href{https://gcn.gsfc.nasa.gov/other/GW190910h.gcn3}{GCN archive}) due to the previous alert (S190910d) which was just a few hours before, in addition to the very large localization.
Only ZTF covered a significant portion of the localization (about 34\% in g/r-band, \citealt{gcn25722}).

\subsection{\textbf{S190923y}}
The candidate S190923y was identified by H1 and L1 at 2019-09-23 12:55:59.646 UTC. So far, only low-latency classification and sky localizations are publicly available \citep{gcn25814}. S190923y is classified with NSBH ($>$68\%) and Terrestrial (32\%) with low latency estimation. The bayestar initial sky localization area gives a 90 \% credible region of 2107 deg$^2$ with a luminosity distance of $438 \pm 133$\,Mpc. Due to the large uncertainty of the sky localization area and the distance luminosity above the completeness of most of the galaxy catalogs (see $\approx$ 17 reports in \href{https://gcn.gsfc.nasa.gov/other/GW190923y.gcn3}{GCN archive}), S190923y has been followed-up by surveys as GRANDMA-TAROT and MASTER in optical bands at $\approx$ 18 mag \citep{gcn25847,gcn25812}.

\subsection{\textbf{S190930t}}

The candidate was identified by L1 at 2019-09-30 14:34:07.685 UTC. So far, only low-latency classification and sky localizations are publicly available \citep{gcn25876}. S190930t is classified with NSBH (74\%) and Terrestrial (26\%). The bayestar initial sky localization area gives a 90\% credible region of 24220 deg$^2$ with a luminosity distance of $108 \pm 38$\,Mpc. A number of the survey instruments, including ATLAS \citep{gcn25922}, MASTER \citep{gcn25712}, and ZTF \cite{gcn25899} covered a significant portion of the localization above $\approx$ 19.5 mag.

\subsection{\textbf{Summary}}

There are a few takeaways from the above.
The first is that dedicated robotic facilities, either in their generic survey mode or performing target of opportunity observations, are present throughout all events. 
Facilities such as TAROT, ZTF, and MASTER, all robotic survey instruments, contributed to kilonova searches for the vast majority of objects. However, we conducted calculation of joint coverage of the sky localization area for two different alerts S190510g and S190814bv with the three networks. The improvement in terms of time spent for exploring a large portion of the skymap is not huge due to the missing coordination of the individual groups. However, this approach might help in terms of having a certain location on the sky re-observed several times which potentially improves the constraints or detection prospects upon further data analysis.
As can be seen from the table, other robotic survey systems also imaged portions of the localizations (for example, with their routine searches for near earth objects), but these serendipitous observations and associated new candidates were not always reported publicly.
This may motivate use of the central reporting databases, if only to assess the level of coverage.
In addition, one notices that, generally, the participation from other systems, at the candidate identification level at least, seemed to have dropped off as the semester went along.


\section{Modeling kilonova and deriving possible limits from observations}
\label{sec:limits}

\subsection{Kilonova modelling}
\label{sec:models}

We will employ three different kilonova models based on \cite{KaMe2017}, \cite{Bul2019}, and \cite{HoNa2019} deriving constraints on possible kilonova lightcurves and their connected ejecta properties. 
With the use of multiple models, we hope to reduce systematic effects. For Model I and Model II, we employ a Gaussian Process Regression based interpolation \citep{DoFa2017} to create a surrogate model for arbitrary ejecta properties (see ~\cite{CoDi2018b,CoDi2018} for further details). The idea of this algorithm is to create interpolated, surrogate models for bolometric lightcurves, photometric lightcurves, or spectral energy distribution in sparse simulation sets typically provided by modeling software. For the photometric lightcurves, in particular, each passband is individually interpolated onto the same time array of 0.1\,days and analyzed separately. To support the interpolation, we perform a singular value decomposition (SVD) of a matrix composed of these lightcurves (separately for each passband); using this, we find eigenvalues and eigenvectors, which we will interpolate across the parameter space. To do so, we use the \texttt{sci-kit learn} \citep{scikit-learn} implementation of Gaussian process regression \citep[GPR,][]{GPRBook}, which is a statistical interpolation method which produces a posterior distribution on a function $f$ given known values of $f$ at a few points in the parameter space.  
Model III is semi-analytic.

\textbf{Model I, [Kasen et al., 2017]}:
For the models presented in \cite{KaMe2017}, each lightcurve depends on the ejecta mass $M_{\rm ej}$, the mass fraction of lanthanides $X_{\rm lan}$, and the ejecta velocity $v_{\rm ej}$. To simplify the analysis, we use a 1-component model which captures the broad features of AT2017gfo as shown in \cite{CoDi2017}, in contrast to the use of a 2-component model \citep{CoDi2018} which improves the fit slightly but doubles the number of free parameters. 
We compute lightcurves consistent with the following prior choices: $-3 \leq \log_{10} (M_{\rm ej}/M_\odot) \leq 0$, $ 0 \leq v_{\rm ej} \leq 0.3$\,$c$. For the ejecta velocity, this covers the range used in the \cite{KaMe2017} simulation set; for the ejecta masses, where the simulation set covers $-3 \leq \log_{10} (M_{\rm ej}/M_\odot) \leq -1$, taking the prior to an ejecta mass of $1 M_\odot$ was chosen for the purpose of upper limits that did not depend on the upper bound. For the lanthanide fraction, we will pin the values to $X_{\rm lan}$ = [ $10^{-9}$, $10^{-5}$, $10^{-4}$, $10^{-3}$, $10^{-2}$, $10^{-1} ]$; note that for ATF2017gfo, assuming the exact same model, a lanthanide fraction of $10^{-3.54}$ 
described the observational data best~\cite{CoDi2018}. \\

\textbf{Model II, [Bulla, 2019]}: 
For the 2-component models presented in \cite{Bul2019}, each lightcurve depends on four parameters: the ejecta mass $M_{\rm ej}$, the temperature at 1~day after the merger $T_0$, the half-opening angle of the lanthanide-rich component $\Phi$ (with $\Phi=0$ and $\Phi=90^\circ$ corresponding to one-component lanthanide-free and lanthanide-rich models, respectively) and the observer viewing angle $\theta_{\rm obs}$ (with $\cos\theta_{\rm obs}=0$ and $\cos\theta_{\rm obs}=1$ corresponding to a system viewed edge-on and face-on, respectively). Unlike \cite{KaMe2017}, models by \cite{Bul2019} do not solve the full radiative transfer equation but rather simulate radiation transport for a given multi-dimensional ejecta morphology adopting parametrized opacities as input. The main advantage over Model I is the possibility to compute viewing-angle dependent observables for self-consistent multi-dimensional geometries in place of combining one-component models with different compositions and thus neglecting the interplay between different components. For this article, we compute lightcurves consistent with $-3\leq \log_{10} (M_{\rm ej}/M_\odot) \leq 0$, $15^\circ\leq \Phi \leq30^\circ$ and $0\leq \cos\theta_{\rm obs}\leq1$,  while the temperature is fixed to the following values: $T_0 = [3000, 5000, 7000, 9000]$~K. 
Note that for ATF2017gfo, $T_0 = 5000$\,K, $\Phi = 30^\circ$, and $\cos\theta_{\rm obs}=0.9$ described the observational data best~\citep{DhBu2019}.
Similar to the \cite{KaMe2017} model, the simulation set covers $-3 \leq \log_{10} (M_{\rm ej}/M_\odot) \leq -1$, and we extend the prior to an ejecta mass of $1 M_\odot$. \\

\textbf{Model III, [Hotokezaka and Nakar 2019]}: 
For the 2-component models presented in \cite{HoNa2019}, the light curves are computed based on the Arnett analytic model \citep{arnett1982} and a black body spectrum with a specific temperature at the photosphere. It assumes spherical ejecta of which the inner part is composed of high-opacity material and the outer part is composed of low-opacity material. In this model, thermalization of gamma-rays and electrons produced by each radioactive decay is taken into account according to their injection energy.
Each light curve depends on $M_{\rm ej}$, the ejecta velocity $v_{\rm ej}$, the dividing velocity between the inner and outer part and the opacity of the 2-components, $\kappa_{\rm low}$ and $\kappa_{\rm high}$. The same prior range for the ejecta mass and velocity as in Model I is used. The model also depends on the lower and upper limit of the velocity distribution, which we set as free parameters within the range of $v_{\text{min}}/v_{\text{ej}} \in [0.1, 1.0]$ and $v_{\text{max}}/v_{\text{ej}} \in [1.0, 2.0]$.

\textbf{Model-independent remarks}: 
Model~I, Model~II, and Model~III use similar nuclear heating rates $\epsilon_\mathrm{nuc}$, in units of ergs per second per gram. Model~I assumes $\epsilon_\mathrm{nuc} = 10^{10}\,t^{-1.3}$~erg~g$^{-1}$~s$^{-1}$ where $t$ is in days \citep{MeMa2010}. Model~II, instead, adopts heating rates from \citet{KoRo2012}, \mbox{$\epsilon_\mathrm{nuc} = \epsilon_0\,\big(\frac{1}{2}-\frac{1}{\pi}\,\mathrm{arctan}\frac{t-t_0}{\sigma}\big)^\alpha\,\big(\frac{\epsilon_\mathrm{th}}{0.5}\big)$}, with $\epsilon_0 = 2\times10^{18}$~erg~g$^{-1}$~s$^{-1}$, $t_0 = 1.3$~s, $\sigma = 0.11$~s, $\alpha = 1.3$ and $\epsilon_\mathrm{th}=0.5$. 
In principle, Model~III computes the radioactive power using the solar r-process abundance pattern with a minimum atomic mass number of $85$. This is however computationally too expensive when sampling over many light curves, so in the analysis presented in Section \ref{sec:ej_mass} we fix the heating rate to the same as Model~II. Although the previous formula provides a better description of nuclear heating rates at short timescales, $t\lesssim10$~s, the agreement between the different rates is excellent at epochs of interest in this study, $t\gtrsim1$~d. 

Within our analysis, we compare the lightcurves to one-sided Gaussian distributions, where we have taken the mean to be the upper limit from the telescope in the given passband and the mean distance from the gravitational-wave skymaps. 
We include a distance variation in our analysis by sampling over a changing ``zeropoint'' in the lightcurves consistent with the distance uncertainty stated in the GW alerts. This is computed by adding a distance modulus consistent with the distance variation from the localizations. While this approach does not account for the exact three-dimensional skymap, it provides representative constraints and limits.

\begin{figure}[t]
 \includegraphics[width=1.1 \columnwidth]{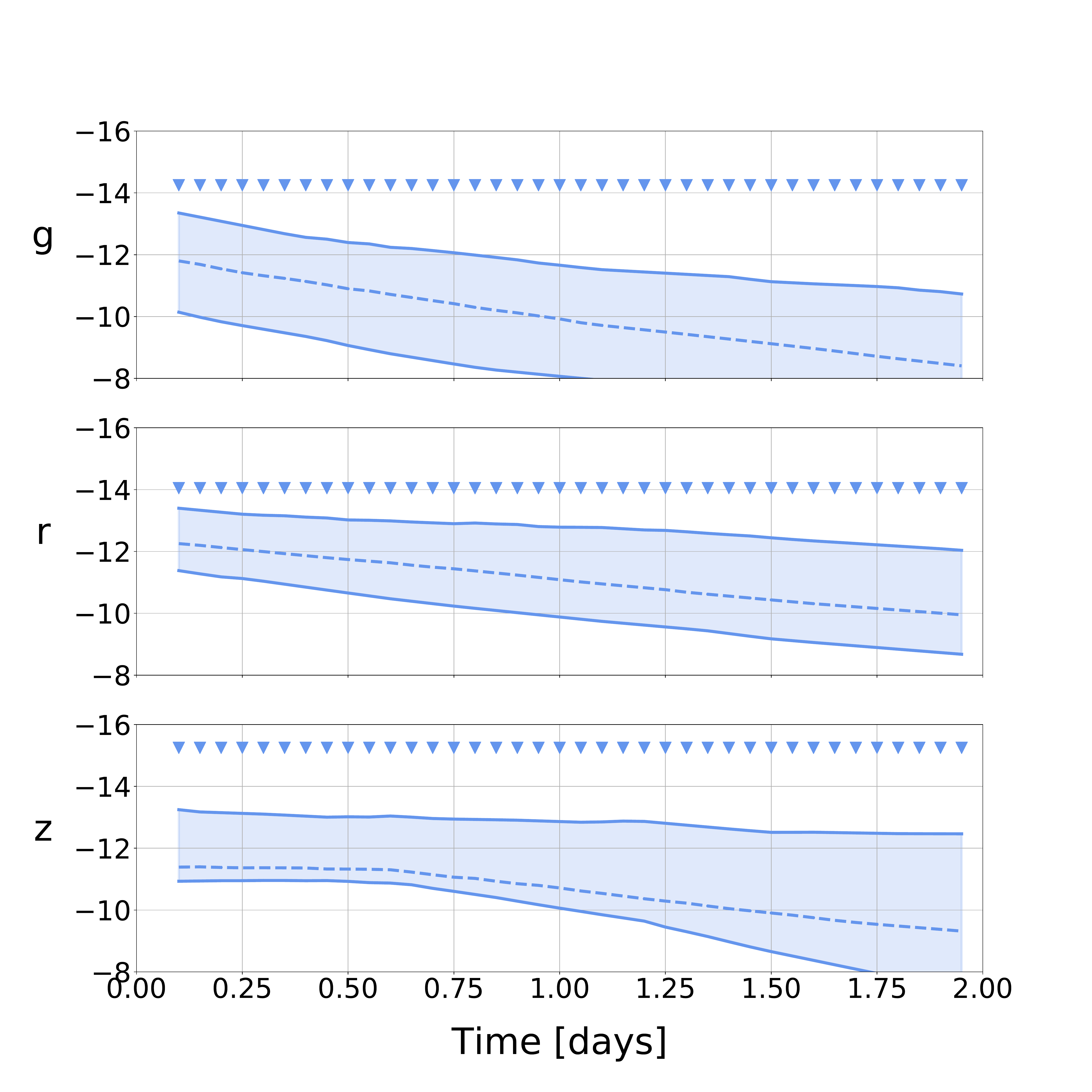}
 \caption{Variety of lightcurves consistent with the Dark Energy Camera based g/r/z limits on S190510g \citep{AnGo2019}, where we show median and 90\% contours for lightcurves based on the \cite{KaMe2017} model.}
 \label{fig:S190510g}
\end{figure}

Figure~\ref{fig:S190510g} gives an example of this approach for the candidate S190510g using the model of \cite{KaMe2017}. It shows the upper limits derived from the Dark Energy Camera in horizontal lines for the three photometric bands $g$, $r$, and $z$. The absolute magnitudes correspond to the mean of the gravitational-wave distance. We also plot an example lightcurve consistent with these constraints. These include the uncertainty in distance sampling. Histograms of the ejecta masses (and other quantities) are made based on these lightcurves, creating the distributions derived in the following analyses.

\subsection{Ejecta mass limits}
\label{sec:ej_mass}

\begin{figure*}[t]
\centering
\textbf{Model I} \hspace{1.6in} \textbf{Model II} \hspace{1.6in} \textbf{Model III} \\
\vspace{0.1in}
\textbf{S190425z} \\
 \includegraphics[width=2.2in,height=1.5in]{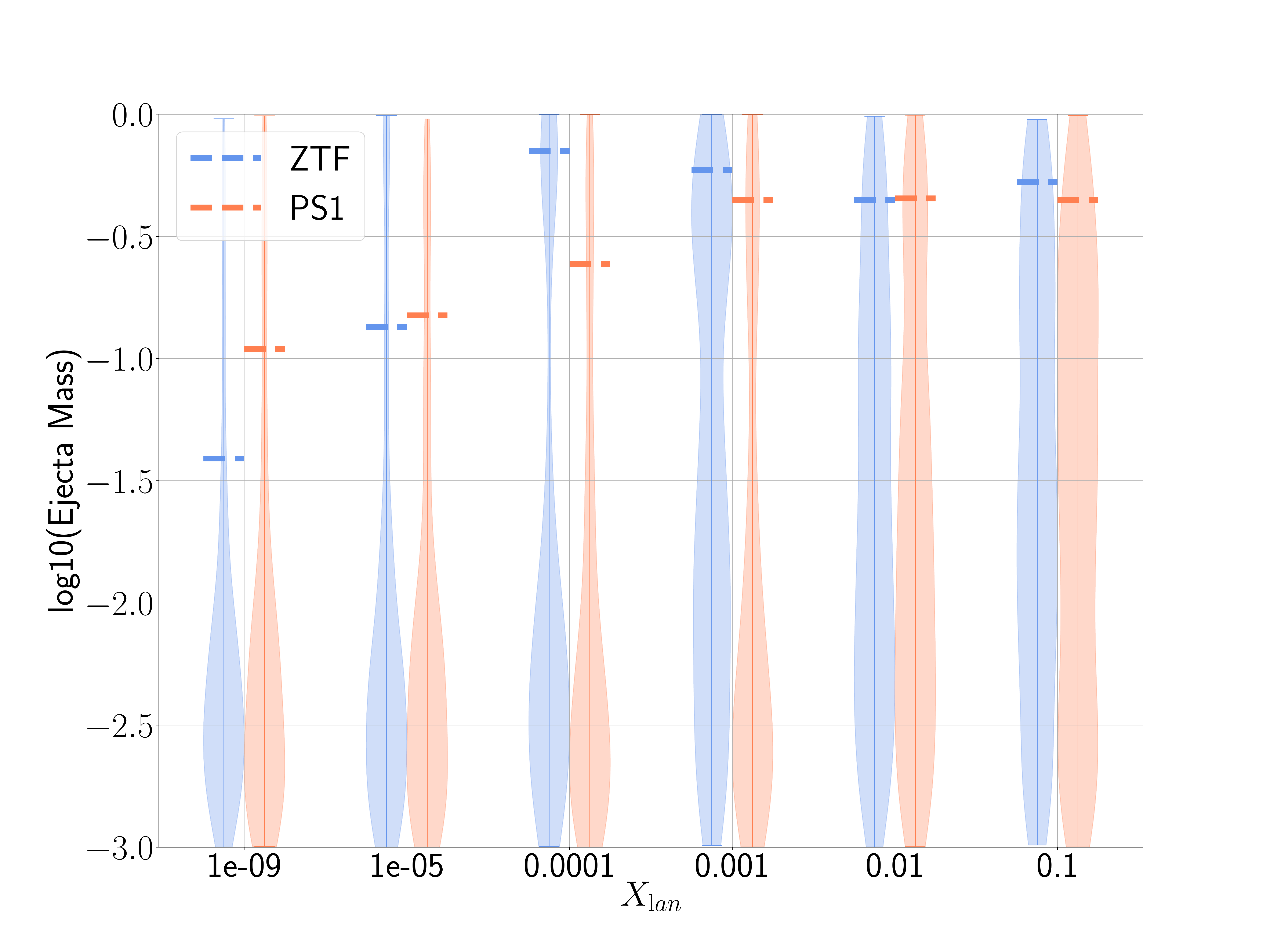}
 \includegraphics[width=2.2in,height=1.5in]{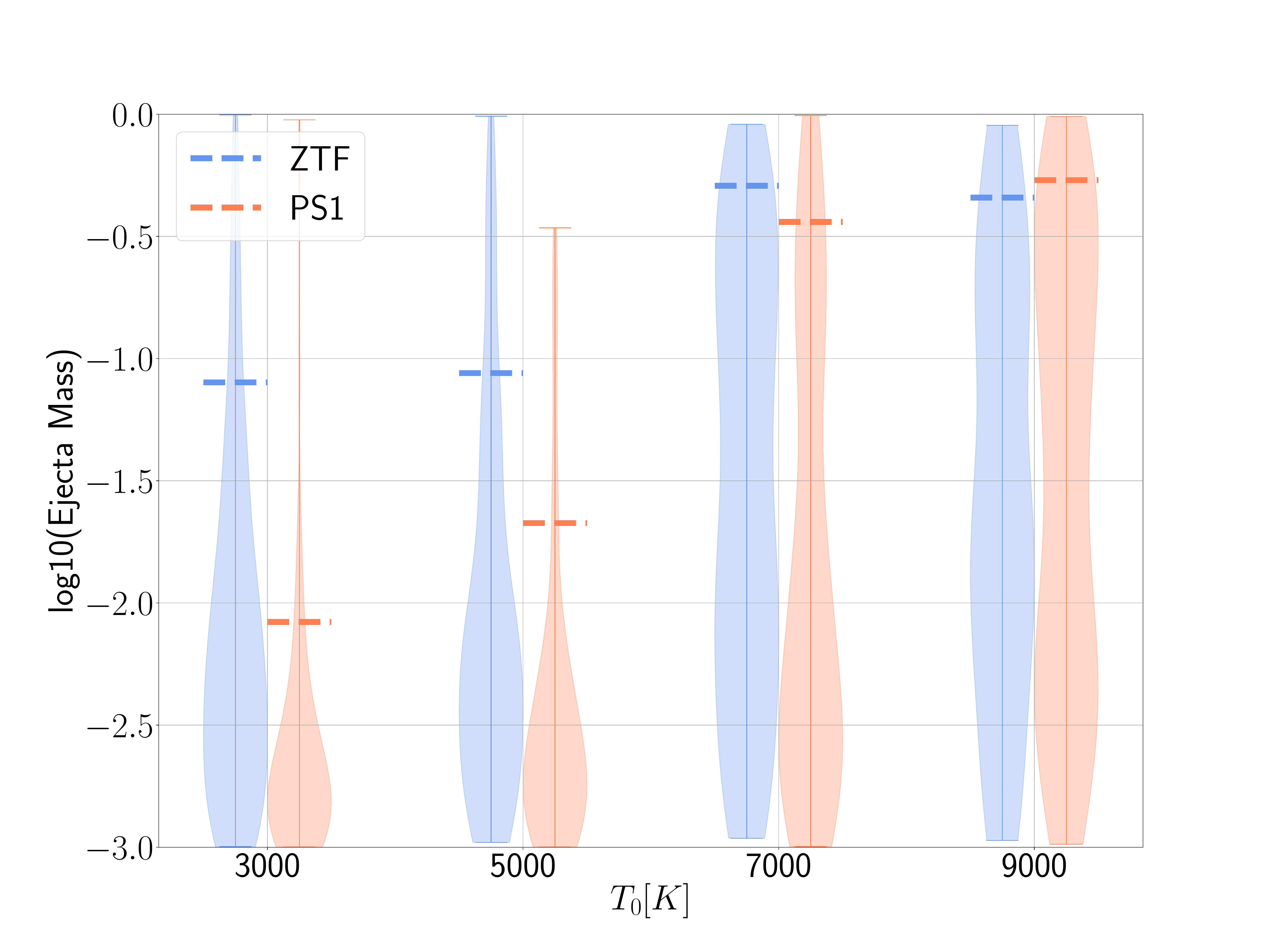}
 \includegraphics[width=2.2in,height=1.5in]{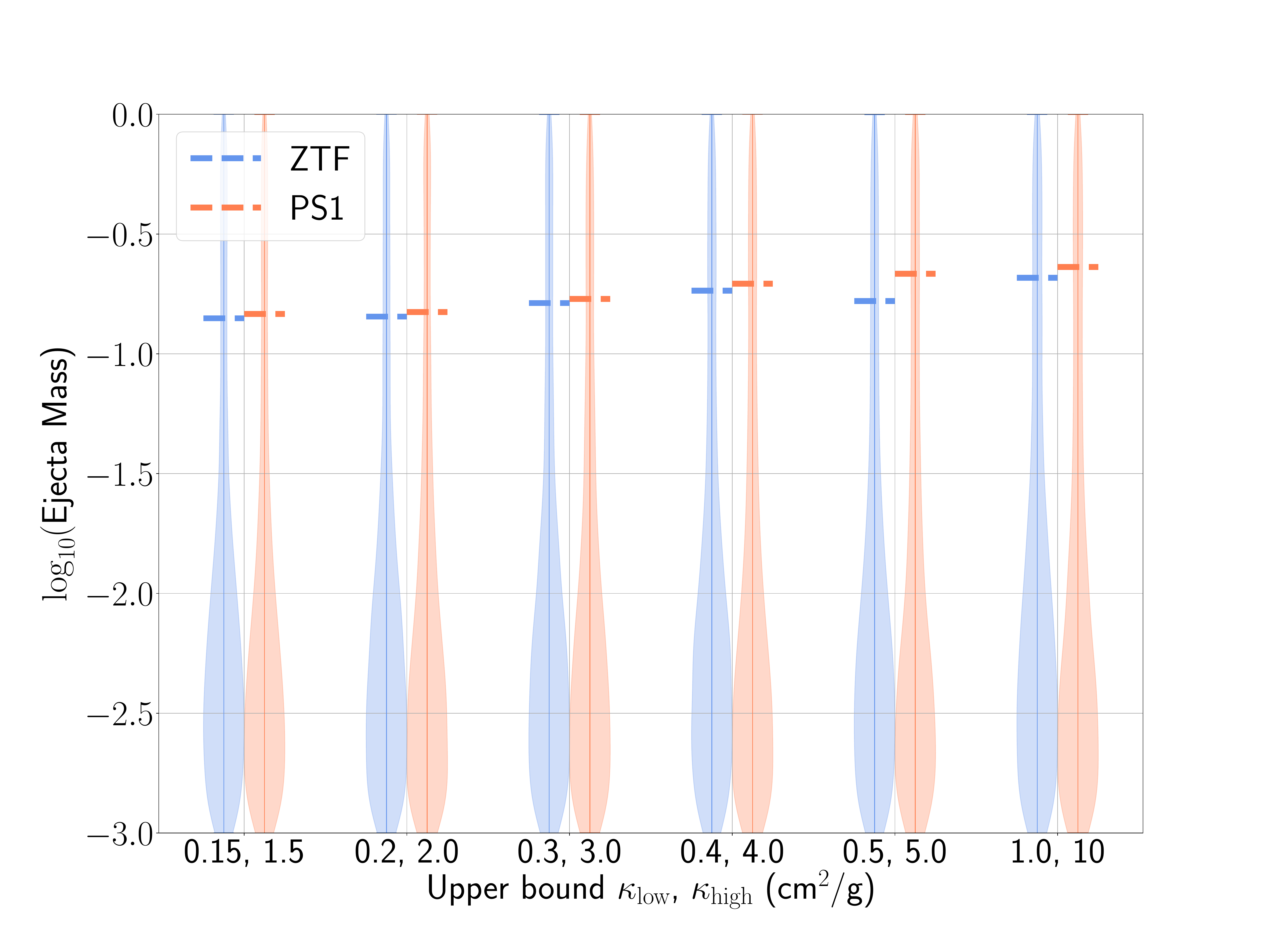}\\
\textbf{S190426c} \\ 
 \includegraphics[width=2.2in,height=1.5in]{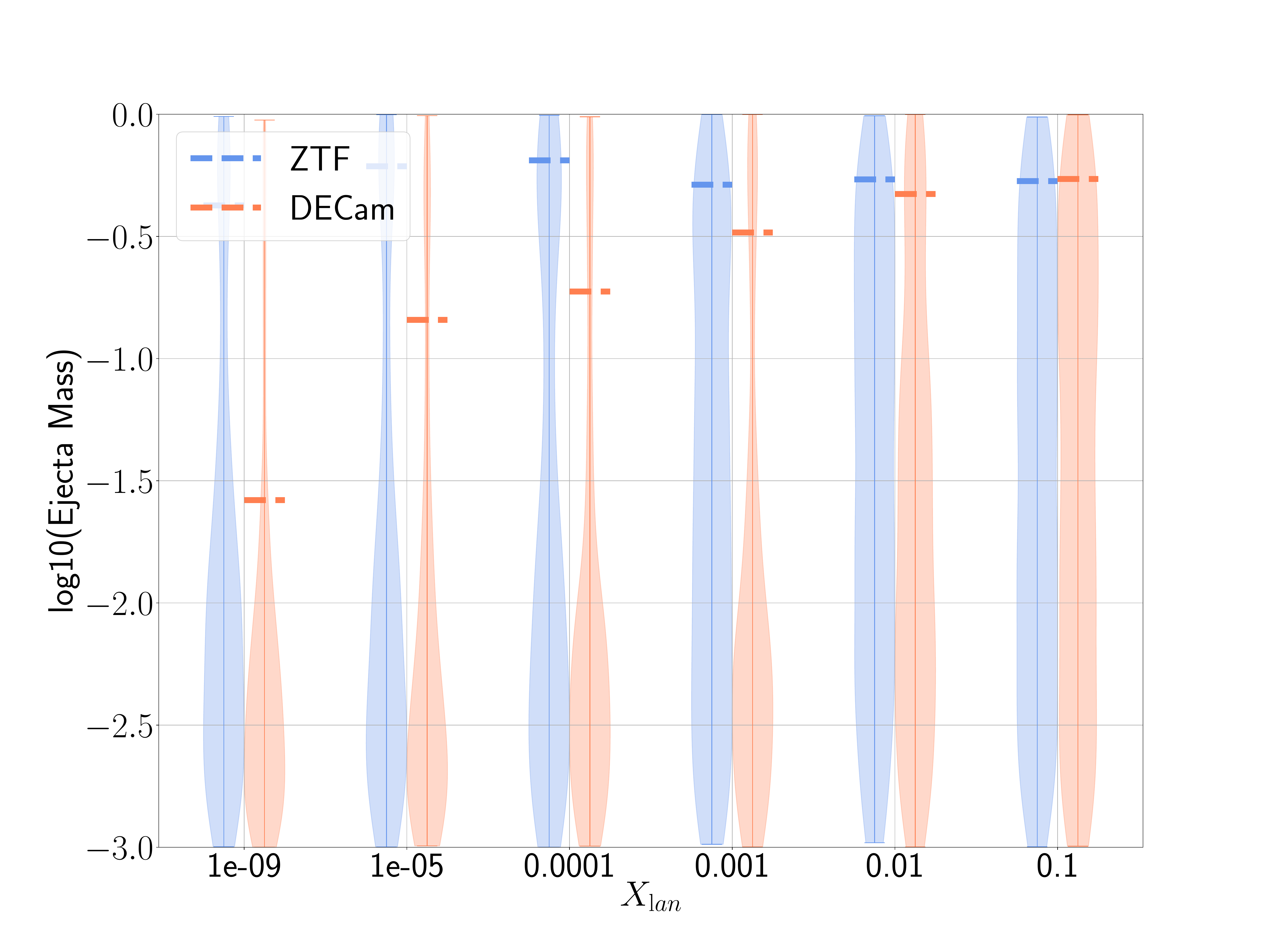}
 \includegraphics[width=2.2in,height=1.5in]{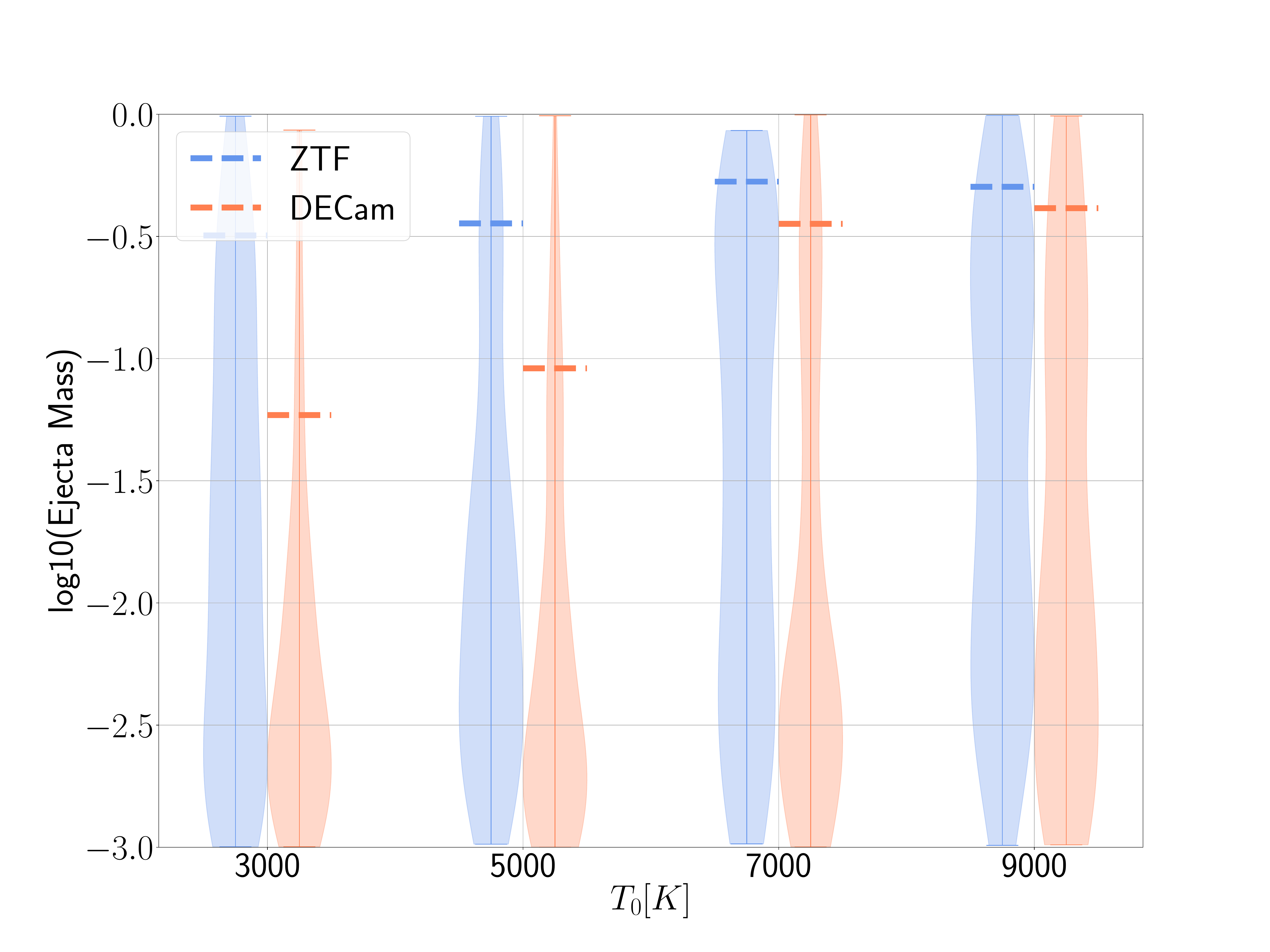}
 \includegraphics[width=2.2in,height=1.5in]{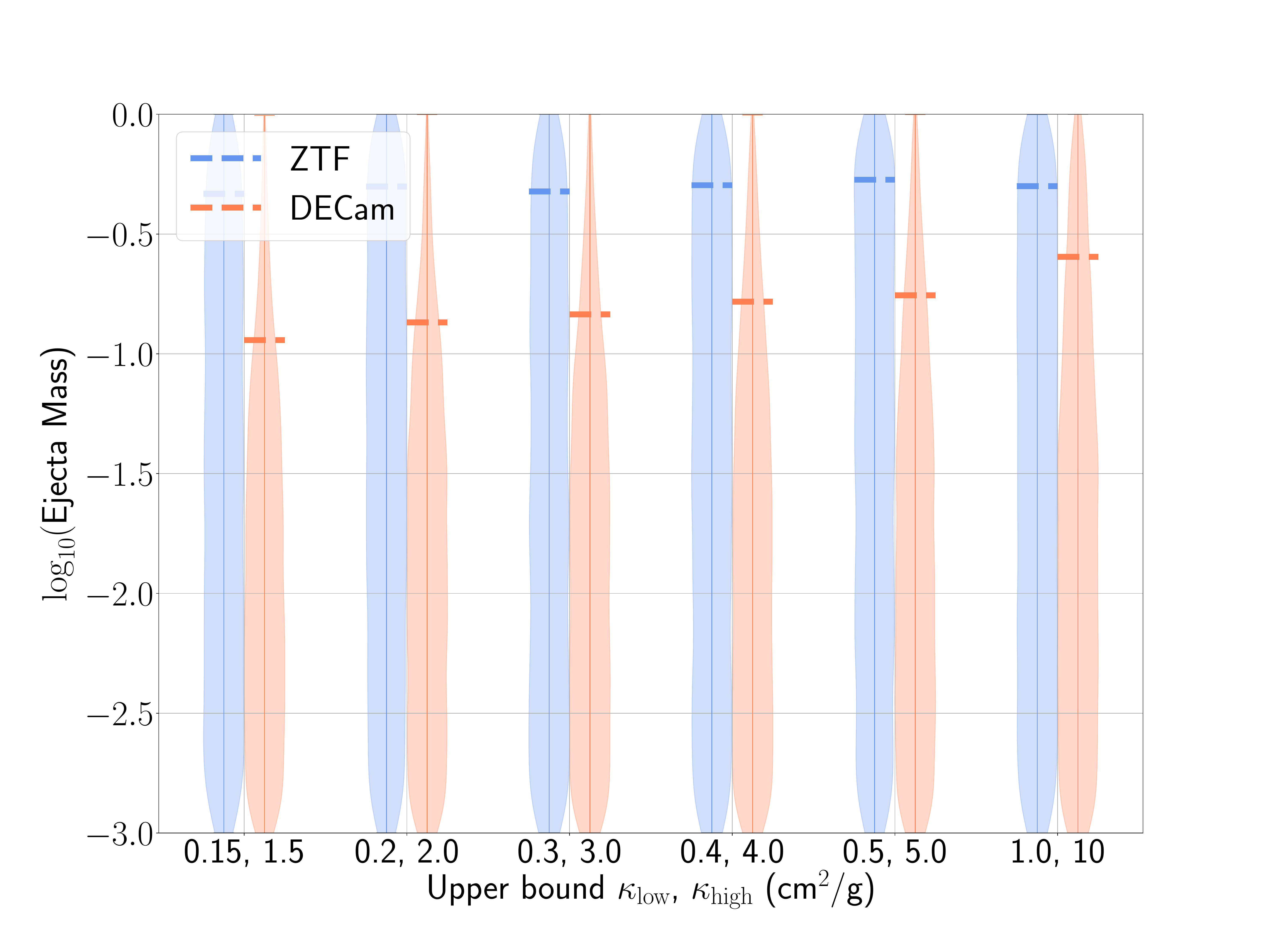}\\
\textbf{S190510g} \\  
 \includegraphics[width=2.2in,height=1.5in]{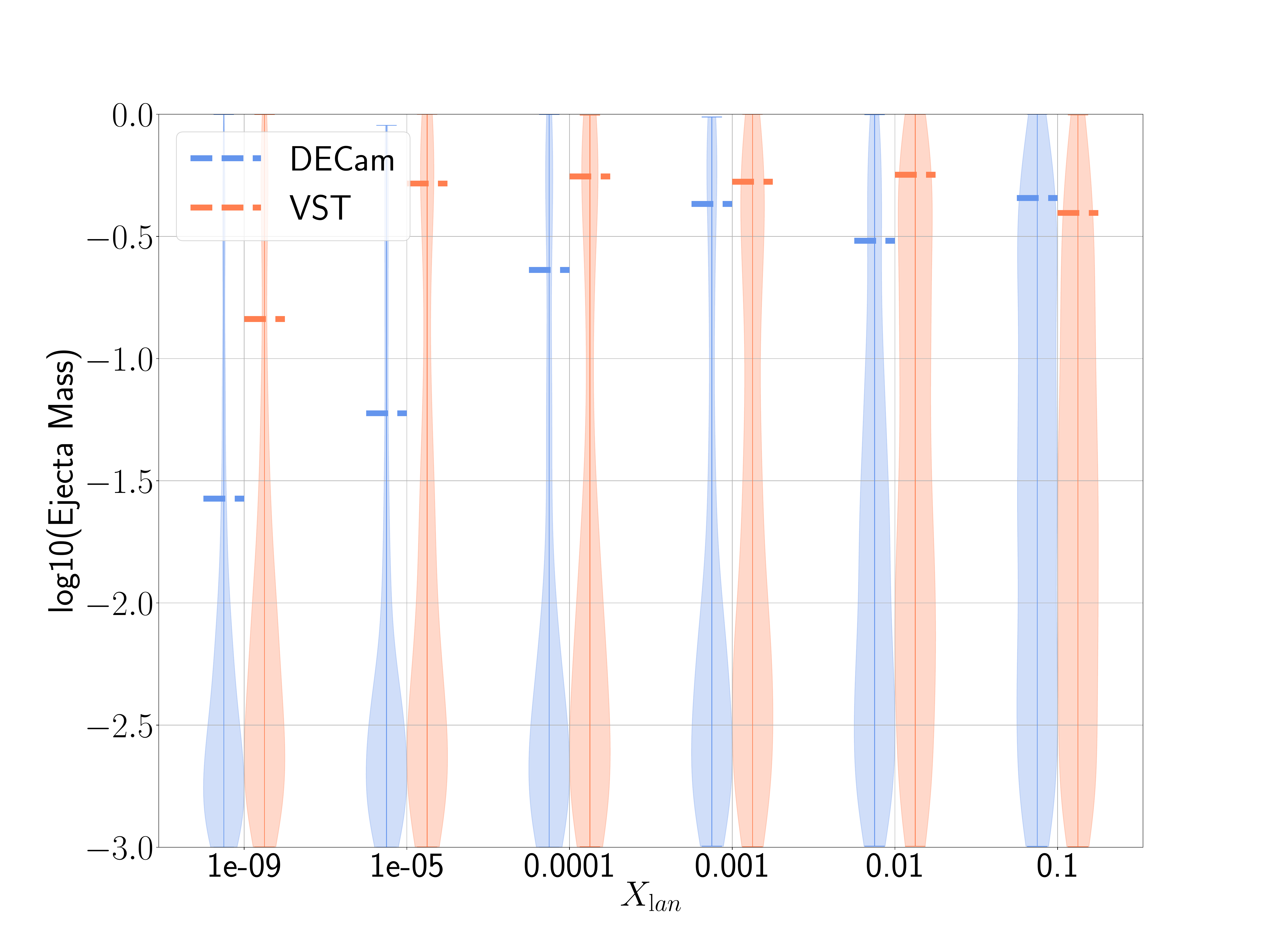}
 \includegraphics[width=2.2in,height=1.5in]{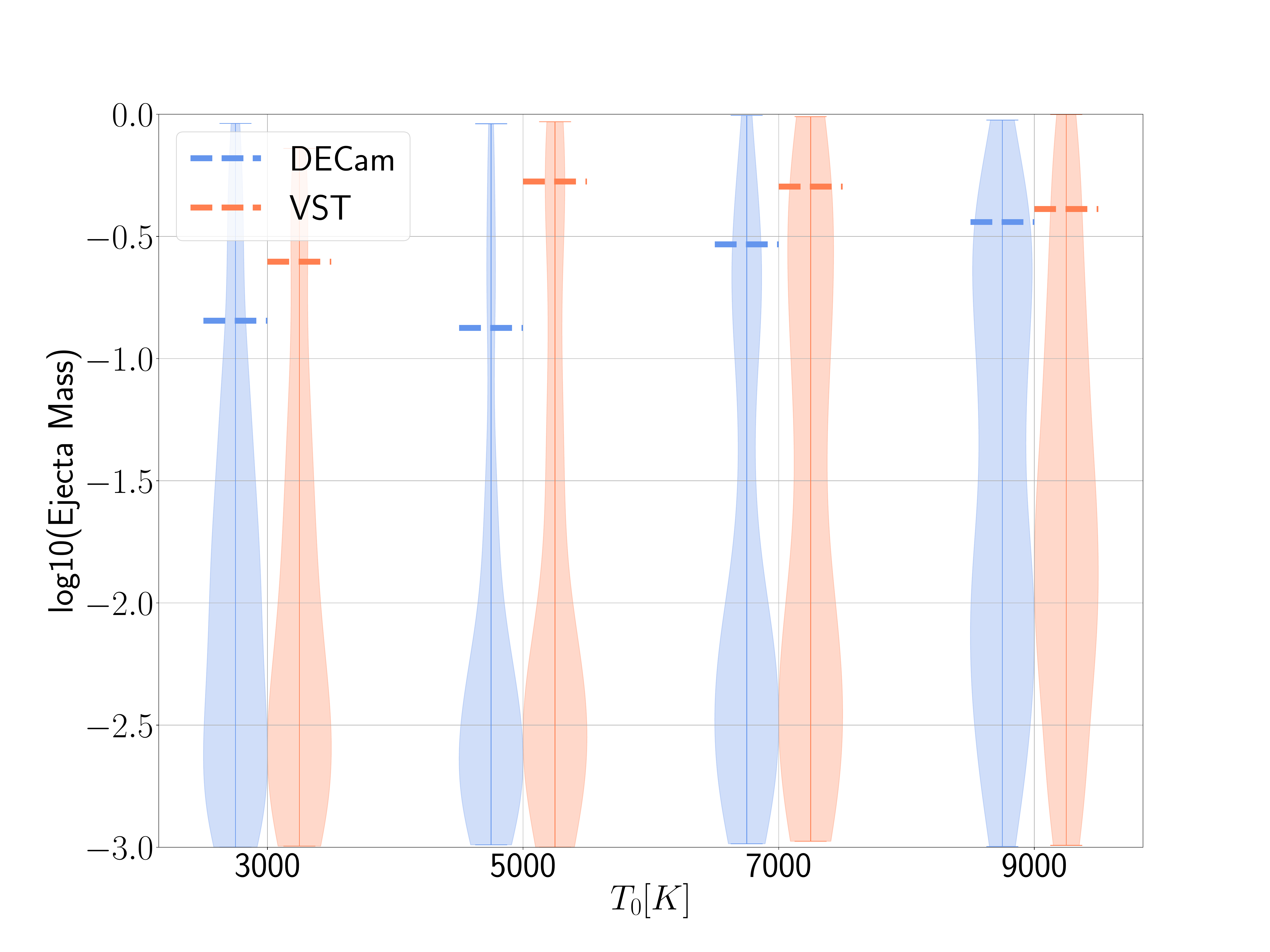}
 \includegraphics[width=2.2in,height=1.5in]{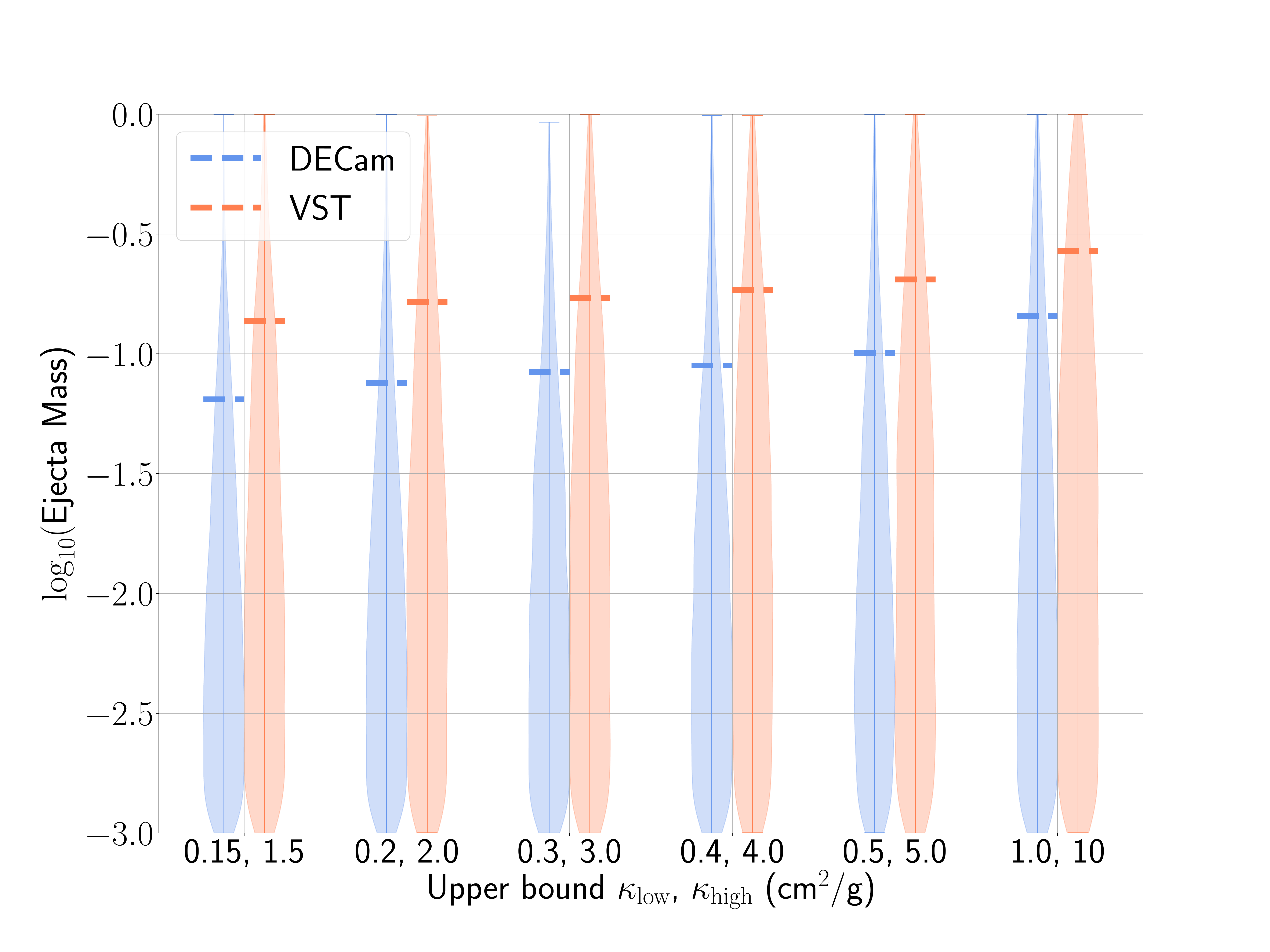}\\
\textbf{S190901ap} \\  
 \includegraphics[width=2.2in,height=1.5in]{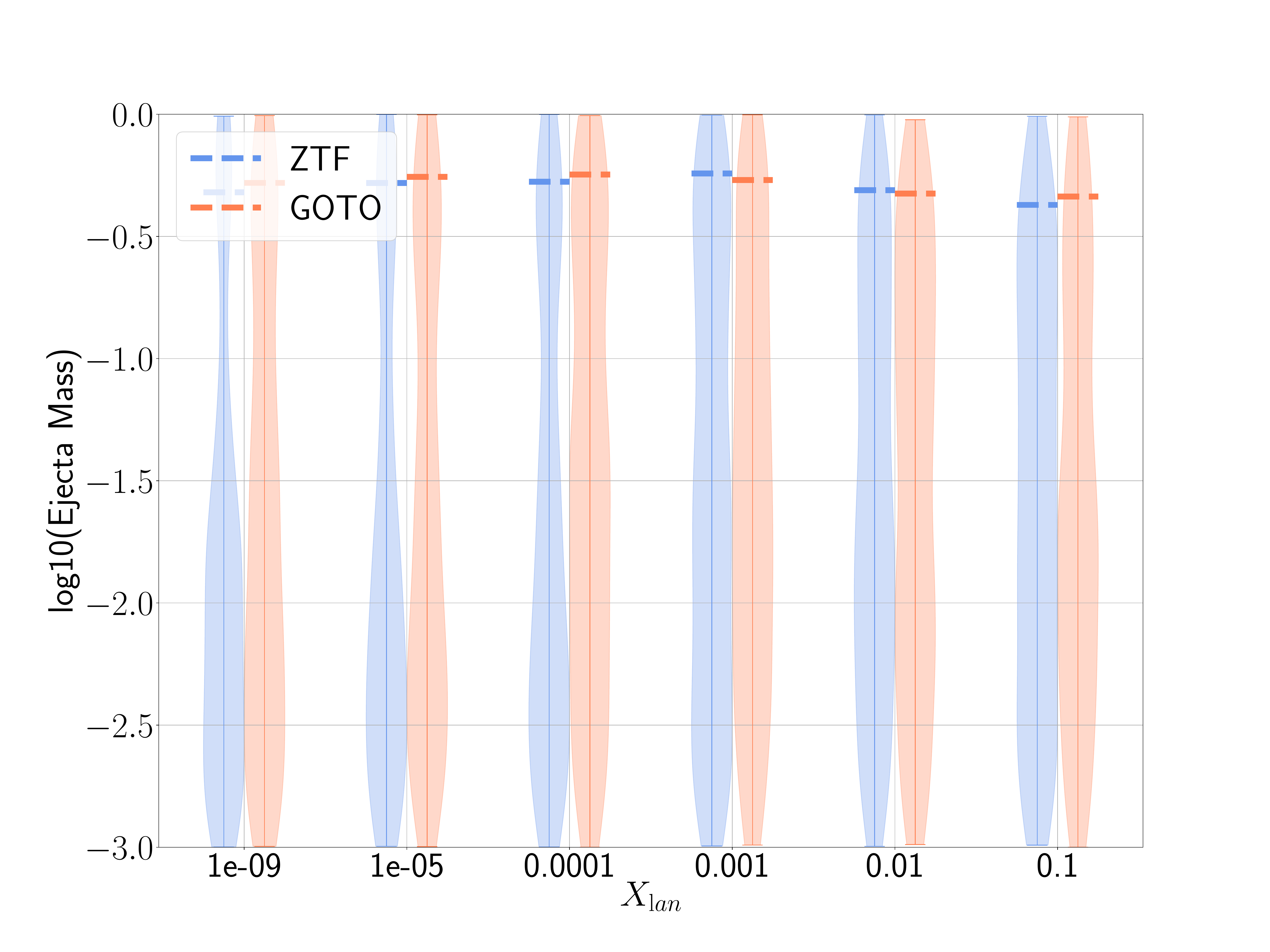}
 \includegraphics[width=2.2in,height=1.5in]{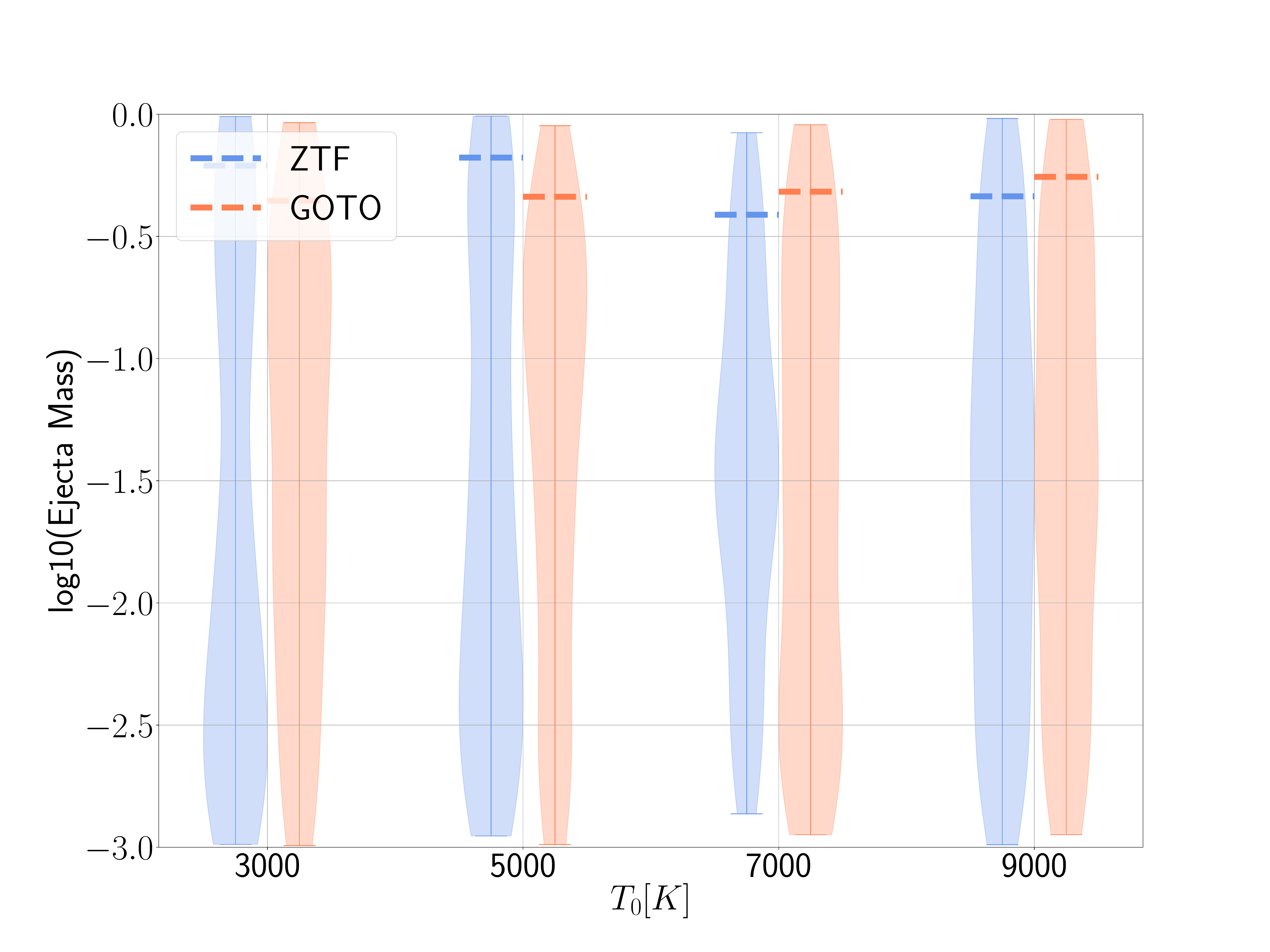}
 \includegraphics[width=2.2in,height=1.5in]{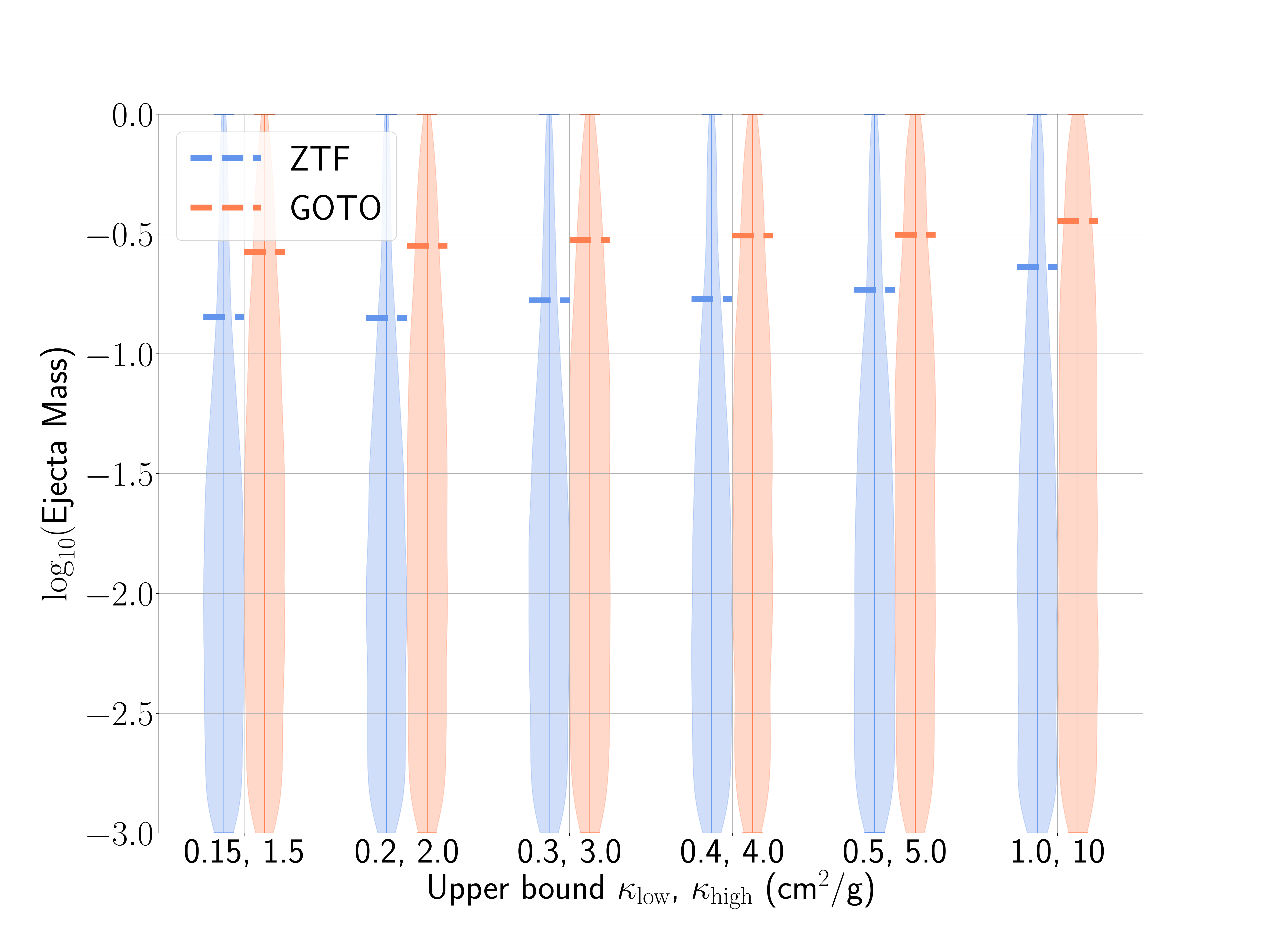}\\
\textbf{S190910h} \\ 
 \includegraphics[width=2.2in,height=1.5in]{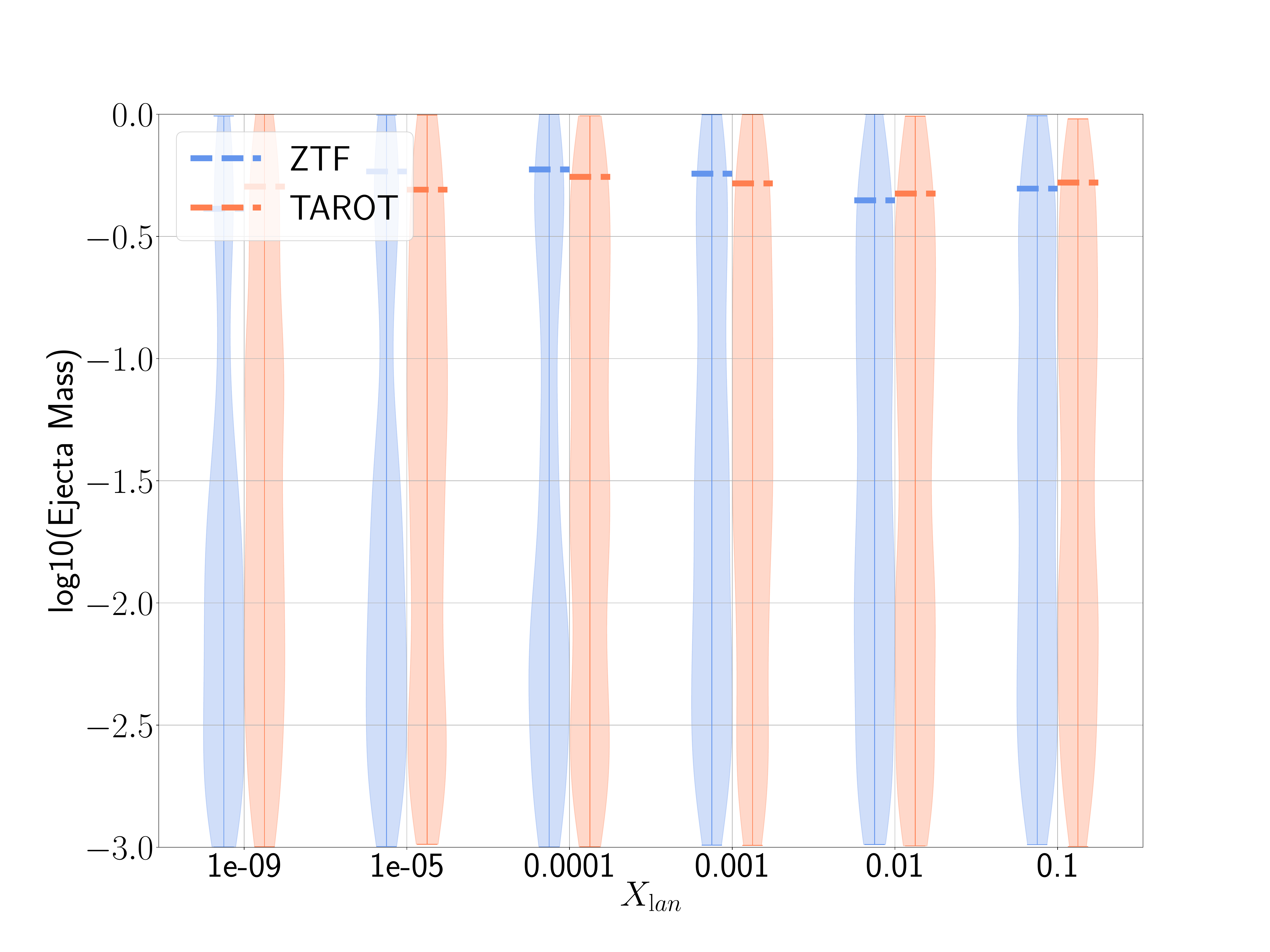}
 \includegraphics[width=2.2in,height=1.5in]{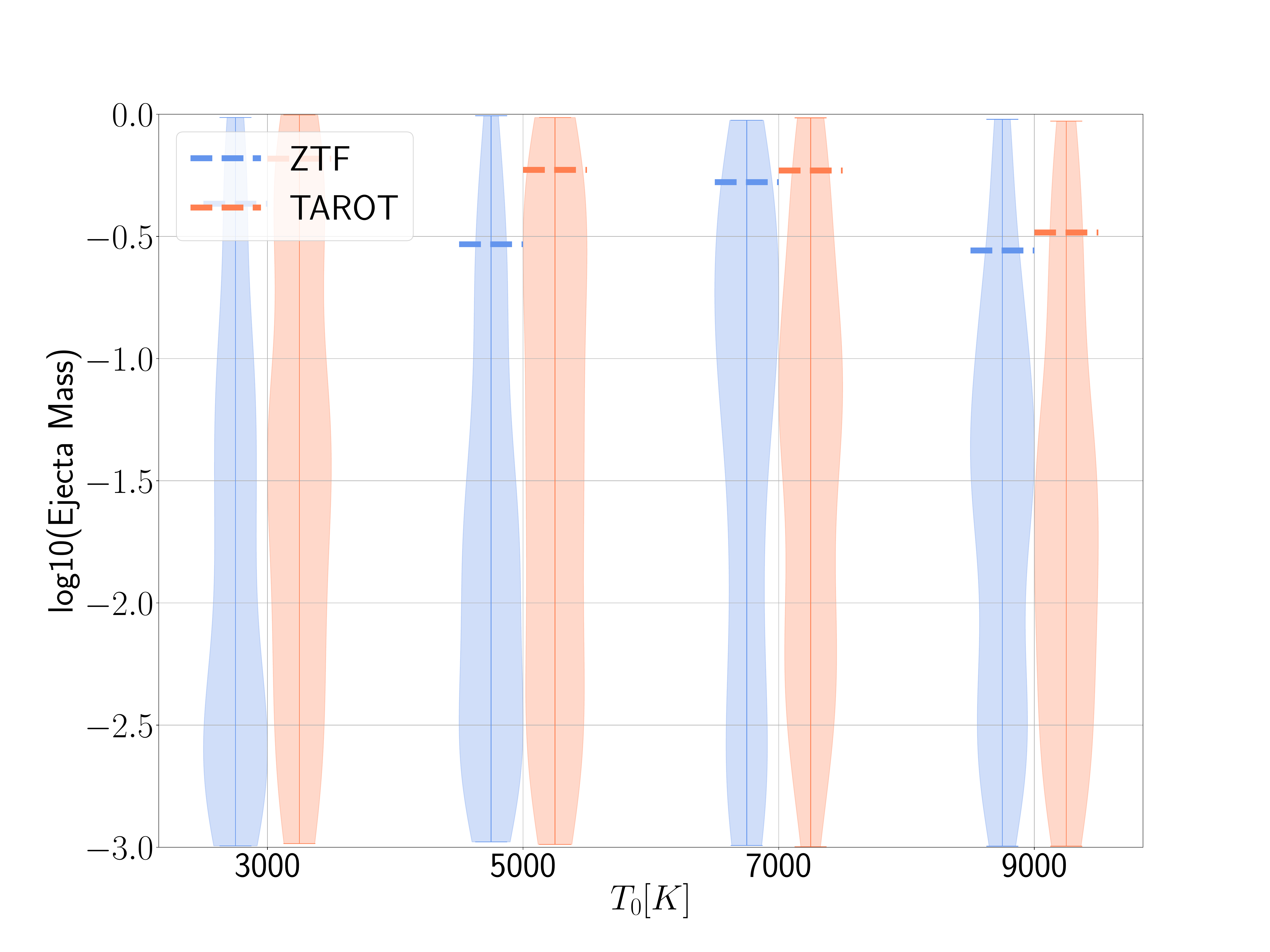}
 \includegraphics[width=2.2in,height=1.5in]{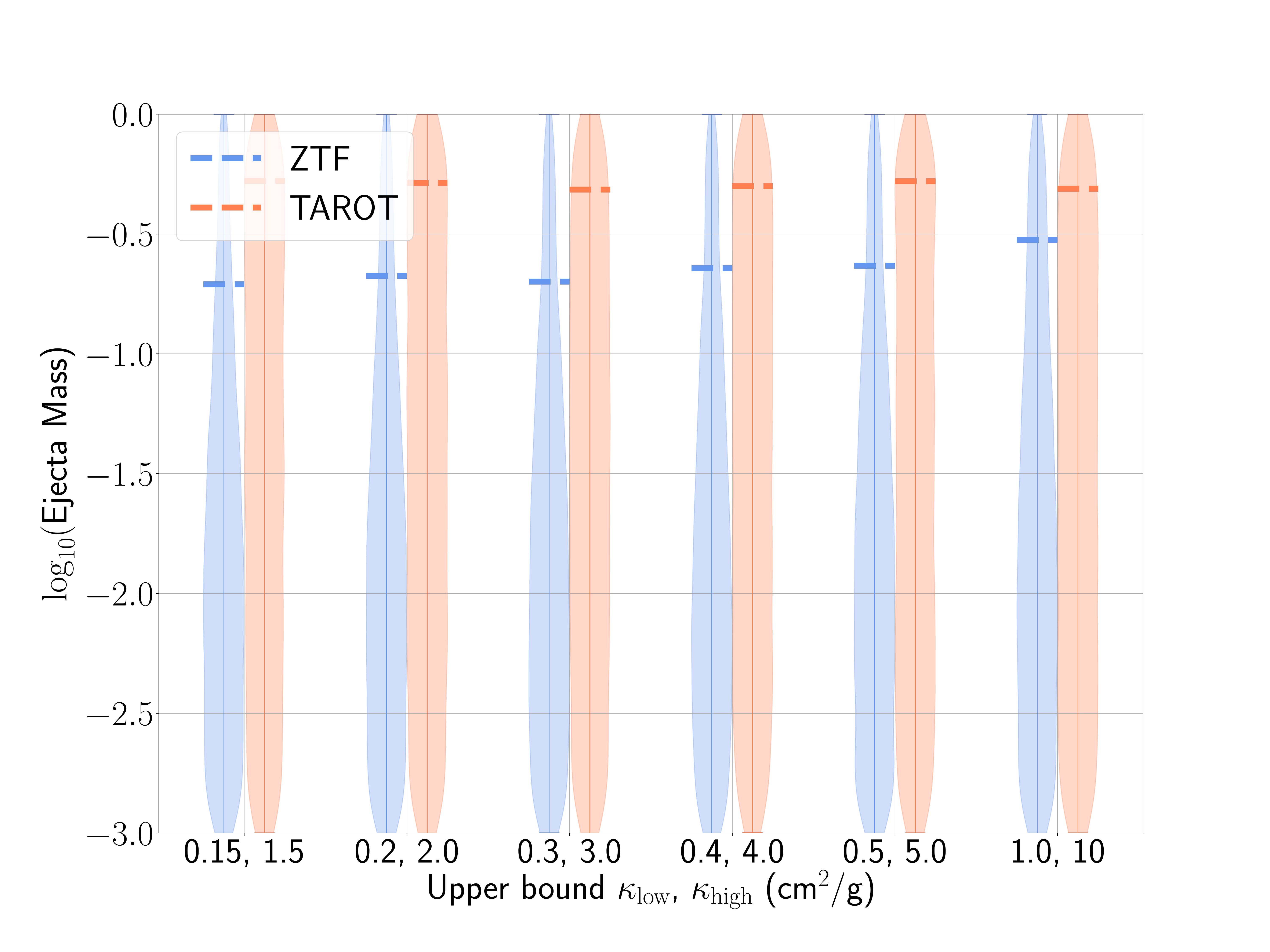}\\ 
 \caption{Probability density for the total ejecta mass for all considered events and all employed lightcurve models. From the top down, the events are S190425z, S190426c, S190510g, S190901ap, and S190910h. From the left to the right, we show constraints as a function of lanthanide fraction for based on the \cite{KaMe2017} model, as a function of temperature for the \cite{Bul2019} model with $\Phi=0^\circ$, and as a function of the opacity of the 2-components, $\kappa_{\rm low}$ and $\kappa_{\rm high}$. For S190425z, we use the ZTF (left, \cite{gcn24191}) and PS1 (right, \cite{gcn24210}) limits. For S190426c, we use the ZTF \citep{gcn24283} and the DECam \citep{gcn24257} limits. For S190510g, we use DECam \citep{AnGo2019} and VST \citep{gcn24484}. For, S190901ap, we use the ZTF \citep{gcn25616} and GOTO \citep{gcn25654} observations. For S190910h, we use the ZTF \citep{gcn25722} and TAROT \citep{gcn25780} observations.}
 \label{fig:violin_constraints}
\end{figure*}

In this section, we provide ejecta mass constraints from comparing different lightcurve models to observational upper limits for S190425z, S190426c, S190510g, S190901ap and S190910h. 
Specifically, we compute ejecta mass constraints for different values of one key quantity for each model: the lanthanide fraction (Model~I), the temperature (Model~II) and the opacities (Model~III). Constraints on the ejecta mass are controlled by the impact of these three different parameters on the predicted kilonova brightness and color. 
Increasing the lanthanide fraction ($X_{\rm lan}$, Model~I) and opacities ($\kappa_{\rm low}$ and $\kappa_{\rm high}$, Model~III) shifts the escaping radiation to longer wavelengths and, thus, leads to the transition from a ``blue'' to a ``red'' kilonova. The impact of the temperature (Model~II) on the brightness and color depends on the epoch since merger. However, at phases when data are most constraining ($\lesssim$~2~d) an increase in temperature results in a shift of the emitted radiation from redder to bluer wavelengths. In particular, moving temperature from $3000$ to $9000$~K produces increasingly fainter kilonovae in both optical and near-infrared bands at these epochs.

Because of the different color predictions, telescopes observing in different regions of the spectrum are associated with different ejecta mass limits. For instance, optical telescopes are generally more constraining to ``blue'' kilonovae that have low lanthanide fractions. \\

\textbf{S190425z:} The top row of figure~\ref{fig:violin_constraints} shows the ejecta mass constraints for S190425z based on observations from ZTF (left, \cite{gcn24191}) and PS1 (right, \cite{gcn24210}).
We mark the $90\%$ confidence with a horizontal dashed line.
In general, the constraints on ejecta mass for the low lanthanide fractions are stronger than available for the ``red kilonovae,'' which are hidden in the redder photometric bands, cf.~Model I. This is a result of using optical telescopes, which cover a large percentage of the sky localization, but are generally more constraining to ``blue'' kilonovae, i.e. those that have low lanthanide fractions.
The $i$-band observations of PS1 lead to stronger constraints on the red side than is possible with ZTF for Model~II, with similar constraints for Model~I and Model~III.
With the higher intrinsic luminosities from Model~II, the constraints in the redder bands from PS1 lead to notable improvements in the constraints.
These constraints are not realized in Model~I and Model~III due to their lower intrinsic luminosities.
We find that the different treatments of the heating rates and radiative transport, yield significantly different ejecta mass constraints than imposed by the effective opacity, temperature, and lanthanide fraction differences, i.e., differences between the three models are larger than within the individual models.
Most notably, Model~II produces, across all considered temperature ranges, the most stringent constraints. 
Consequently, while Model~I and Model~III only disfavor (in the most optimistic scenarios) ejecta masses $M_{\rm ej} \lesssim 0.1M_\odot$, which is very hard to achieve for a BNS merger, Model II places upper bounds on the ejecta mass of $M_{\rm ej} \lesssim 0.03M_\odot$ for temperatures at or below $5000K$.

\textbf{S190426c:} The second row of figure~\ref{fig:violin_constraints} shows the ejecta mass constraints for S190426c based on the observations from ZTF \citep{gcn24283} and the DECam \citep{gcn24257}. Despite the smaller sky area requiring coverage and therefore generally deeper exposures, the larger distance to this object leads to limits that are worse than for the first event. However, for a number of parameter combinations, we find that ejecta masses above $\sim 0.1M_\odot$ are ruled out based on the DECam observations. Furthermore, as for S190425z, one obtains tighter constraints for blue kilonova (low lanthanide fractions and opacities) for Model~I and Model~III, and for redder kilonovae in Model~II.

\textbf{S190510g:} The third row of figure~\ref{fig:violin_constraints} shows the ejecta mass constraints for S190510g based on observations from DECam \citep{AnGo2019} and VST \citep{gcn24484}. The relative improvement of sensitivity between ZTF and DECam offsets the relative difference in distance estimates, yielding very similar ejecta mass constraints between the two binary neutron star coalescence candidates, i.e., S190510g and S190425z. The inclusion of the three bands, g-, r-, and z-band observations with DECam produces measurable constraints in both the blue and red bands; for example, with Model I, $M_{\rm ej}$ is $\lesssim 0.025M_\odot$ for the lowest lanthanide fractions.

\textbf{S190901ap:} The fourth row of figure~\ref{fig:violin_constraints} shows the ejecta mass constraints for S190901ap based on observations from ZTF \citep{gcn25616} and GOTO \citep{gcn25654}. 
Due to the large sky localization covering more than 10,000 deg$^2$, there was relatively minimal EM follow-up investigation. The larger distance to this potential BHNS system results in the shallowest constraints on ejecta mass for all considered candidates.

\textbf{S190910h:}
The final row of figure~\ref{fig:violin_constraints} shows the ejecta mass constraints for S190910h based on observations from ZTF \citep{gcn25722} and TAROT \citep{gcn25780}. 
Due to the large sky localization covering more than 20,000 deg$^2$, there was relatively minimal EM follow-up investigation, and therefore, similar to the event above, there were essentially no constraints.

\textbf{Summary}:
Considering the five individual constraints, we find that S190425z and S190426c provide overall the tightest constraints for a BNS and BHNS candidate, respectively. 
However, our analysis shows that even for these events, no constraints can be obtained with Model~III or for Model~I in case for ejecta with high lanthanide fractions. These loose constraints are mainly caused by the large distance to the individual candidate events, which are generally several times further away than GW170817. 
Considering the results obtained from Model~II, we will describe in the next section how potential ejecta mass constraints lead to constraints on the binary properties of BNS and BHNS candidates. 
However, we want to emphasize that there are large systematic differences between the lightcurve models and that the entire sky area provided by LIGO and Virgo has not been covered for all triggers. Thus, the following analysis should be rather interpreted as a proof of principle. 

\section{Constraining the binary parameters}
\label{sec:binary_property}

Within this section, we present as a proof of principle possible constraints for the binary properties of the BNS candidate S190425z and the BHNS candidate S190426c (under the assumption that the source location was covered within the EM follow-up campaign). We focus on the results of Model~II with a fixed temperature of $5000\rm K$\footnote{With the chosen temperature of $5000\rm K$ the predictions of Model~II agree best with AT2017gfo~\citep{DhBu2019}. Thus, this temperature choice seems best suited for our analysis.}. This leads to a maximum total ejecta masses of $0.03M_\odot$ for S190425z and $0.09M_\odot$ for S190426c.

\subsection{The binary neutron star candidate S190425z}

To ensure that the ejected material is massive enough to trigger a bright EM counterpart, the final remnant should not collapse promptly to a black hole (BH) after the merger. As mentioned in the introduction, prompt collapse formation depends dominantly on the total mass of the binary. As shown in~\cite{Bauswein:2013jpa} the total mass of the binary $M$ has to be below a characteristic threshold mass:
\begin{equation}
    M_{\rm thr} = \left( 2.380 - 3.606 \frac{M_{\rm TOV}}{R_{1.6M_\odot}}  \right) M_{\rm TOV} \label{eq:Mth_Bauswein}
\end{equation}
with $M_{\rm TOV}$ being the maximum supported mass for a spherical NS and $R_{1.6M_\odot}$ the radius of a $1.6M_\odot$ NS. Recently, the threshold mass estimate was updated by~\cite{Koppel:2019pys} incorporating a non-linear dependence on the maximum allowed compactness and \cite{Agathos:2019sah} derived a prompt-collapse threshold estimate based on new numerical relativity simulations, mainly publicly available at \texttt{http://www.computational-relativity.org} \citep{Dietrich:2018phi}. For our rough estimates presented here, we will use, for simplicity, the criterion given in \cite{Bauswein:2013jpa}.  

\begin{figure}[t]
 \includegraphics[width=\columnwidth]{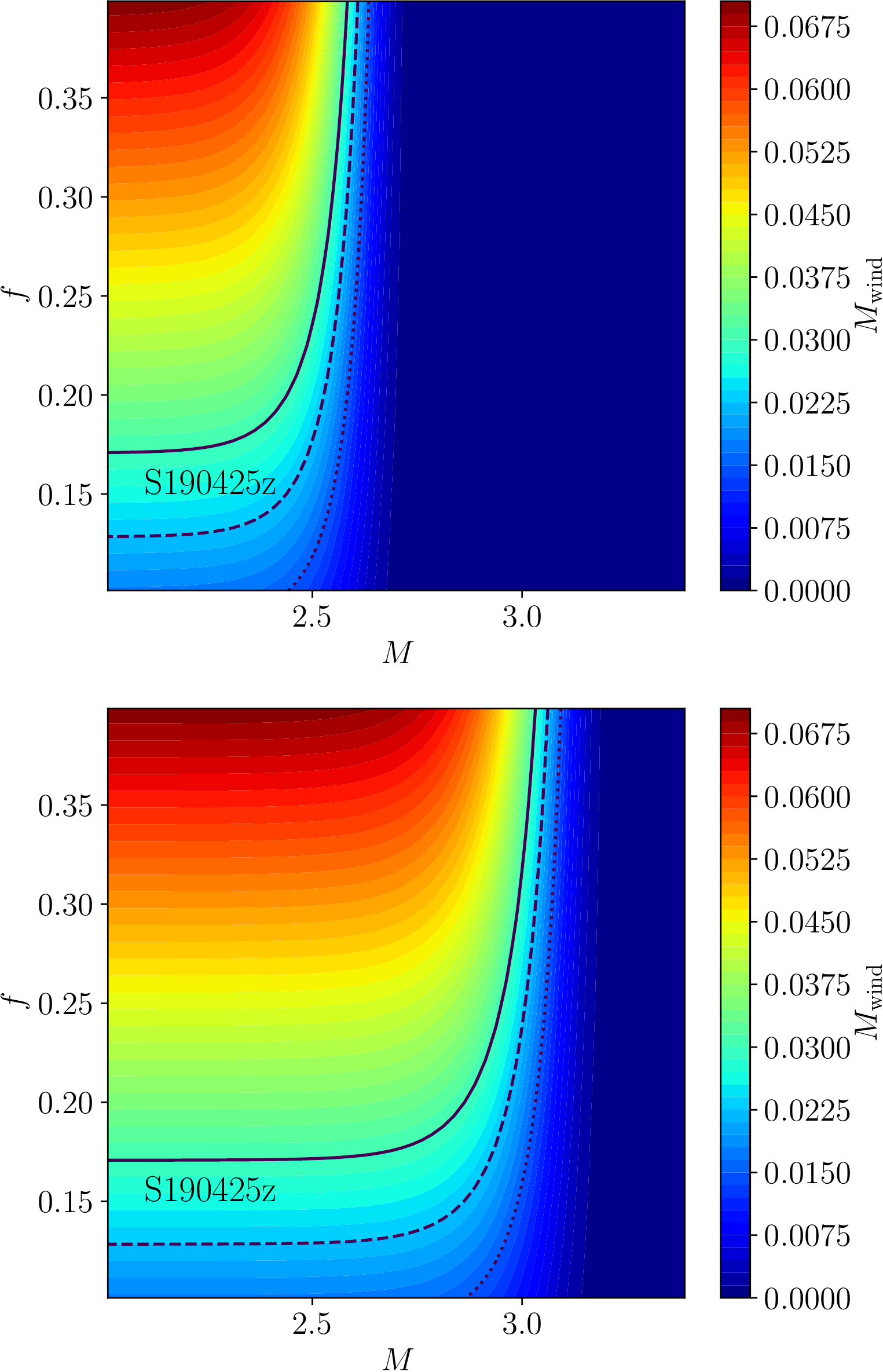}
 \caption{Disk wind ejecta as a function of the conversion factor $f$ and the total mass of the binary $M$. We include the upper bounds from S190425z using Model II for $5000\rm K$ assuming that the disk wind accounts for the entire ejecta mass (solid line), $75\%$ (dashed line) or for $50\%$ of the total ejecta (dotted line). The top panel assumes an EOS with a maximum TOV mass of $2.07M_\odot$ and a radius of $R_{1.6M_\odot} = 11.1\rm km$, while the bottom panel uses $M_{\rm TOV}=2.30M_\odot$ and $R_{1.6M_\odot} = 13.9\rm km$.}
 \label{fig:mass_BNS}
\end{figure}

While for close GW events it would be a valid assumption that all configurations without an EM counterpart have masses above the prompt threshold mass, $M>M_{\rm thr}$, this assumption does not hold for systems with distances much larger than the one for GW170817, e.g., for S190425z.
In general, the total ejecta mass, for which our previous analysis provided some upper limits, is related to the debris disk mass formed after the merger; here, we use the disk mass estimate presented in \cite{CoDi2018b}, where $M_{\rm disk}$ was a function on $M/M_{\rm thr}$:
\begin{align}
\log_{10} &\left( m_{\rm disk} \left[ M_{\rm tot}/M_{\rm thr} \right]\right) = \nonumber \\ 
& \max \left(-3,\ a  \left(1 + b \tanh \left[ \frac{c - M_{\rm tot}/M_{\rm thr}}{d} \right] \right) \right)
\label{eq:mdisk_fit}
\end{align}
with the fitting parameters $a,b,c,d$; see  \cite{CoDi2018b}. We emphasize that this estimate was based on a suite of numerical relativity simulations for equal-mass or near equal-mass systems, high mass ratio systems might lead to more massive disks~\cite{Kiuchi:2019lls}
The mass of the disk wind is then $M_{\rm wind} = f M_{\rm disk}$ with the unknown conversion factor $f$. This efficiency parameter remains very uncertain~\citep{Fernandez:2014cna,Siegel:2017jug,Fernandez:2018kax,Christie:2019lim} and we will vary it for our BNS analysis, $f\in[0.1,0.4].$\footnote{Existing 3D simulations, which seed the accretion disk with a purely toroidal or purely poloidal magenetic field, fall at the high end of that interval, $f\sim [0.3-0.4]$. We conservatively allow for lower values of $f$ to account for the possibility that about half of that ejecta is produced at early times, in magnetically-driven winds that appear to depend on the strength and stucture of the magnetic field and may still disappear for the small-scale turbulent magnetic fields that are most likely created in a neutron star merger.}
Since a fraction of the ejecta will also be released dynamically during the merger, not all of the total ejecta comes from disk winds. As an indication, we present the disk wind estimate in figure~\ref{fig:mass_BNS} assuming $100\%$ of the total ejecta mass for S190425z are connected to the wind ejecta (solid black line), $75\%$ of the total mass is assigned to winds (dashed line), and half of the total ejecta comes from disk wind ejecta (dotted line). 

The two panels in figure~\ref{fig:mass_BNS} refer to different choices of the maximum TOV-mass $2.07M_\odot$ for the top and $2.30M_\odot$ for the bottom panel. These values are motivated by the recent observation of J0740+6620~\citep{Cromartie:2019kug} and the upper bound on the maximum mass following from GW170817, e.g.~\citep{MaMe2017,ReMo2017,Shibata:2019ctb}. In addition, we assume a radius $R_{1.6M_\odot}$ of $11.1\rm km$ in the top and $13.9\rm km$ in the bottom panel, as derived in \cite{CoDi2018b}.These combinations of $M_{\rm TOV}$ and $R_{1.6M_\odot}$ include the most extreme scenarios in terms of stiff and soft EOSs, and, thus, provide boundaries for our analysis. 
Considering the scenario for a very soft EOS, we find that the total mass of S190425z lies presumably above $2.40 M_\odot$ if the efficiency factor if about $20\%$. Contrary for an efficiency factor of $20\%$ and a very stiff EOS, the total mass of S190425z would presumably be $2.9M\odot$. 

\subsection{The black hole - neutron star candidate S190426c}

\begin{figure*}[t]
 \includegraphics[width=\textwidth]{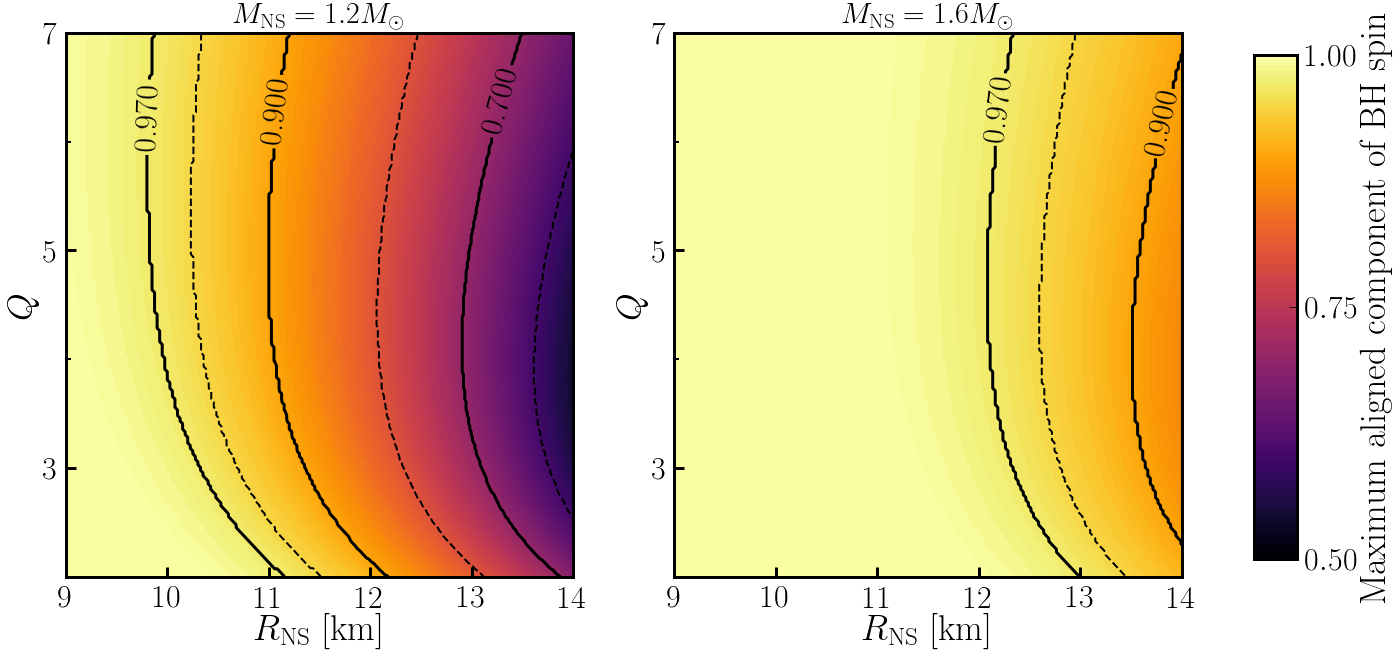}
 \caption{Maximum aligned component of the black hole spin as a function of neutron star radius and binary mass ratio for S190426c, if that event was a BHNS merger \textit{within} the region covered by follow-up observations. We show results for $M_{\rm NS}=1.2M_\odot$ (left) and $M_{\rm NS}=1.6M_\odot$, and require $M_{\rm ej}<0.09M_\odot$ (Model II for $5000\rm K$). Dashes lines correspond to maximum spins of $0.6,0.8,0.95$, and solid lines to maximum spins of $0.7,0.9,0.97$. Only a small part of the parameter space (large aligned spin, low mass neutron stars) can possibly be ruled out be observations. The constraints for more massive stars ($M_{\rm NS}=1.6M_\odot$) are not reliable since the analytical models have not been tested for spins greater than $\sim 0.9$.}
 \label{fig:spin-BHNS}
\end{figure*}

Similarly for BHNS systems, the absence of an observed kilonova constrains the initial parameters of the binary. As for the BNS case, the outflows from BHNS mergers can be divided into the dynamical ejecta, which is produced at the time of merger and typically lanthanide-rich~\citep{Deaton:2013sla,Foucart:2014nda,Kyutoku:2017voj}, and magnetically-driven or neutrino-driven disk winds produced in the $1-10$ seconds following the merger, which have a more uncertain composition~\citep{Fernandez:2014cna,Just:2014fka,Siegel:2017jug,Fernandez:2018kax}. The dynamical ejecta for neutron stars within the range of parameters used in numerical simulations so far is well modeled by the fit of~\cite{Kawaguchi:2016ana}. Extrapolating that fit to more compact stars, however, leads to unphysical results (i.e. an increase in the ejected mass for more compact stars). Here, we use the modified formula
\begin{small}
\begin{equation}
\frac{M_{\rm dyn}}{M^b_{\rm NS}} =
\max \bigg(0,a_1 Q^{n_1} (1-2C_{\rm NS}) -a_2 Q^{n_2}\frac{c^2 R_{\rm ISCO}}{GM_{\rm BH}}+ a_3\bigg)
\end{equation}
\end{small}
with $Q=M_{\rm BH}/M_{\rm NS}$, $M^b_{\rm NS}$ the baryon mass of the neutron star, $C_{\rm NS}=GM_{\rm NS}/(R_{\rm NS}c^2)$ its compactness, $R_{\rm ISCO}$ the radius of the innermost stable circular orbit around the black hole, and $a_1=0.007116, a_2=0.001436, a_3=-0.0276, n_1=0.8636, n_2=1.6840$ (see Kr\"uger et al., in prep, for a more detailed discussion). Note that $R_{\rm ISCO}$ is computed for circular orbits around a black hole of dimensionless spin $\chi_{\rm eff}=\chi_\parallel$, with $\chi_{\parallel}$ the component of the black hole spin aligned with the orbital angular momentum of the binary. As a result, the ejected mass has a strong dependence in the aligned component of the black hole spin.
The total mass in the bound accretion disk surrounding the remnant black hole can be estimated by subtracting $M_{\rm dyn}$ from the total amount of mass remaining outside of the black hole after merger $M_{\rm out}$. We compute $M_{\rm out}$ following the fit to numerical results provided in~\cite{Foucart:2018rjc}. Similarly to $M_{\rm dyn}$, $M_{\rm out}$ depends on the mass ratio of the system, the compactness of the neutron star, and the aligned component of the black hole spin. The mass in the disk winds is then $M_{\rm wind}=f (M_{\rm out}-M_{\rm dyn})$. Since the BHNS case contains already a larger number of free parameters, we fix the conversion factor to $f\gtrsim 0.15$~\citep{Fernandez:2014cna,Siegel:2017jug,Fernandez:2018kax,Christie:2019lim}.

If S190426c was a BHNS merger within the region of the sky observed by ZTF and DECAM, and we assume the constraints obtained with Model II at $5000\rm K$, we argued that $M_{\rm ej}=M_{\rm dyn}+M_{\rm wind}$ has to be less than $0.09M_\odot$. Practically, this can be converted into a constrain excluding part of the 3-dimensional parameter space of $(Q,C_{\rm NS},\chi_{\parallel})$. Figure~\ref{fig:spin-BHNS} visualizes this constraint as a maximum allowed value for the component of the dimensionless black hole spin {\it aligned with the orbital angular momentum} of the binary, as a function of neutron star size and binary mass ratio. We see that with this upper bound, the constraints on the parameter space of BHNS binaries are fairly weak: only large aligned black hole spins combined with low-mass stars and relatively stiff equations of state can possibly be ruled out.

\section{Summary}
\label{sec:summary}

We have presented an overview of the extensive searches for EM transients 
associated with a number of GW event triggers within the first half of the third observing run of Advanced LIGO and Advanced Virgo. 
Assuming that the individual sources were located in the covered sky region of the follow-up observations, we use three different kilonova models to derive possible upper limits on the ejecta mass compatible with the non-observation of EM signals for S190425z, S190426c, S190510g, S190901ap, and S190910h. Possibly informative constraints are obtained for S190425z and S190426c with the model of~\cite{Bul2019}. However, systematic uncertainties between different kilonova models are large and currently the dominating source of error in our analysis. 

Based on our results, we computed potential lower limits on the total mass of S190425z from the non-existence of EM counterparts and find that it should have a total mass above $2.5M_\odot$ if we assume a soft and $2.9M_\odot$ if we assume a stiff EOS. 
Similarly, assuming that S190426c originated from a BHNS merger, we find that the non-observation of a kilonova could rule out large aligned black hole spins combined with low-mass stars (for stiff EOSs). 

Our simple analysis shows that even without direct GW information, beyond the provided skymap and classification probability, source properties can be constrained\footnote{While we used \texttt{HasRemnant} to downselect the events, we did not rely on its results for the analysis.}. More importantly, inverting our approach, one sees that a fast estimation of the total mass can potentially be used to classify if potential GW candidates will cause bright EM counterparts. A similar approach has been recently outlined in~\cite{Margalit:2019dpi}. 

In general, the limits derived on the ejecta mass for the events in the first six months of O3 are not striking, which shows that one should be striving to take deeper observations, perhaps at the cost of a smaller sky coverage. 
Assuming that AT2017gfo is representative, ``interesting'' limits are $\sim\,0.05\,M_\odot$, giving a ballpark limit to strive for.
Those observations are most important at low latency, i.e., at times when kilonovae are brightest.
In addition to adding and/or employing guiding to take longer observations, it might motivate the creation and use of stacking pipelines for survey facilities, for which this may be atypical. \\

\acknowledgments

Michael Coughlin is supported by the David and Ellen Lee Postdoctoral Fellowship at the California Institute of Technology. 
Tim Dietrich acknowledges support by the European Union's Horizon 2020 research and innovation program under grant agreement No 749145, BNSmergers. 
Sarah Antier is supported by the CNES Postdoctoral Fellowship at Laboratoire Astroparticle et Cosmologie. MB acknowledges support from the G.R.E.A.T research environment funded by the Swedish National Science Foundation.
Francois Foucart gratefully acknowledges support from NASA through grant 80NSSC18K0565 and from the NSF through grant PHY-1806278.
The lightcurve fitting / upperlimits code used here is available at: 
\url{https://github.com/mcoughlin/gwemlightcurves}.

\bibliographystyle{mnras}
\bibliography{references}

\appendix

\FloatBarrier

\begin{deluxetable*}{cccccccc}
\tablenum{2}
\setlength{\tabcolsep}{0pt}
\tablecaption{Reports of the observations by various teams of the sky localization area of gravitational-wave alerts of the possible BNS candidates S190425z, S190510g, S190901ap, S190910h and S190930t. Teams that employed ``galaxy targeting'' during their follow-up are not mentioned here. In the case where numbers were not reported or provided upon request, we recomputed some of them; if this was not possible, we add $-$.}
\label{tab:TableobsBNS}
\tablewidth{0pt}
\tablehead{ \begin{scriptsize} Telescope \end{scriptsize} & \begin{scriptsize} Filter \end{scriptsize} & \begin{scriptsize} Limit mag  \end{scriptsize}& \begin{scriptsize}  Delay aft. GW  \end{scriptsize} & \begin{scriptsize}  Duration \end{scriptsize} & \multicolumn{2}{c}{\begin{scriptsize}  GW sky localization area \end{scriptsize}}  & \begin{scriptsize}  reference \end{scriptsize} \\  &  &  & \begin{scriptsize}  (h) \end{scriptsize} &\begin{scriptsize}  (h) \end{scriptsize} & \begin{scriptsize} name \end{scriptsize} & \begin{scriptsize} coverage (\%)  \end{scriptsize}& }
\startdata
\multicolumn{8}{c}{\textbf{S190425z}} \\
\begin{scriptsize} ATLAS    \end{scriptsize} & 
\begin{scriptsize} o-band \end{scriptsize}   & 
\begin{scriptsize} $19.5$   \end{scriptsize} & 
\begin{scriptsize} $0.8$  \end{scriptsize}   & 
\begin{scriptsize} $6.2$  \end{scriptsize}   & 
\begin{scriptsize} bayestar ini  \end{scriptsize} & 
\begin{scriptsize} $37$   \end{scriptsize} & 
\begin{scriptsize} \citet{gcn24197} \end{scriptsize} \\
\begin{scriptsize} CNEOST  \end{scriptsize} & 
\begin{scriptsize} clear \end{scriptsize}   & 
\begin{scriptsize} $\approx 20$\end{scriptsize} &
\begin{scriptsize}  $27.5$   \end{scriptsize}   & 
\begin{scriptsize} $4.8$ \end{scriptsize}  & 
\begin{scriptsize}bayestar ini \end{scriptsize} & 
\begin{scriptsize}$10$ \end{scriptsize} & 
\begin{scriptsize} \citet{gcn24285} \end{scriptsize} \\
\begin{scriptsize} GOTO \end{scriptsize}  & 
\begin{scriptsize}L-band \end{scriptsize} & 
\begin{scriptsize} $20.5$  \end{scriptsize} & 
\begin{scriptsize} $11.7$    \end{scriptsize} &
\begin{scriptsize} $8.9$  \end{scriptsize} & 
\begin{scriptsize}bayestar ini \end{scriptsize} &
\begin{scriptsize} $30$  \end{scriptsize} &
\begin{scriptsize} \citet{gcn24224} \end{scriptsize} \\
\begin{scriptsize} GRANDMA-TAROT \end{scriptsize}   & 
\begin{scriptsize}clear\end{scriptsize}    & 
\begin{scriptsize} $17.5$  \end{scriptsize}& 
\begin{scriptsize} $6.7$    \end{scriptsize}  & 
\begin{scriptsize} $<63$ \end{scriptsize} & 
\begin{scriptsize} LALInference  \end{scriptsize}&
\begin{scriptsize}  $3$  \end{scriptsize}  & 
\begin{scriptsize}\citet{gcn24227}  \end{scriptsize}\\
\begin{scriptsize} GROWTH-Gattini-IR \end{scriptsize} & 
\begin{scriptsize}J-band \end{scriptsize}  & 
\begin{scriptsize} $15.5$   \end{scriptsize}    & 
\begin{scriptsize} $1.0$  \end{scriptsize}      & 
\begin{scriptsize} $27.8$  \end{scriptsize} &
\begin{scriptsize} LALInference  \end{scriptsize}& 
\begin{scriptsize}$19$   \end{scriptsize}& 
\begin{scriptsize} \citet{gcn24187}  \end{scriptsize}\\
\begin{scriptsize} MASTER-network  \end{scriptsize} & 
\begin{scriptsize}clear  \end{scriptsize}  & 
\begin{scriptsize} $\approx 18.5$ \end{scriptsize} & 
\begin{scriptsize} $\approx 0.0$     \end{scriptsize}   & 
\begin{scriptsize}$144$  \end{scriptsize}  &
\begin{scriptsize} bayestar ini  \end{scriptsize}& 
\begin{scriptsize}$37$  \end{scriptsize}  & 
\begin{scriptsize}\citet{gcn24167}  \end{scriptsize} \\
\begin{scriptsize} Pan-STARRS  \end{scriptsize} & 
\begin{scriptsize}i-band \end{scriptsize}  &
\begin{scriptsize}  $21.5$    \end{scriptsize}    & 
\begin{scriptsize} $1.3$      \end{scriptsize} & 
\begin{scriptsize} $<19$   \end{scriptsize}& 
\begin{scriptsize}bayestar ini  \end{scriptsize}& 
\begin{scriptsize}$28$ \end{scriptsize}  & 
\begin{scriptsize}\citet{gcn24210}  \end{scriptsize}\\
\begin{scriptsize} SAGUARO  \end{scriptsize} & 
\begin{scriptsize}g-band \end{scriptsize}  & 
\begin{scriptsize} $\approx 21$\end{scriptsize}  &
\begin{scriptsize} $1.3$    \end{scriptsize}    &
\begin{scriptsize}$1.3$   \end{scriptsize} & 
\begin{scriptsize}bayestar ini \end{scriptsize} & 
\begin{scriptsize}$3$  \end{scriptsize}  & 
\begin{scriptsize}\citet{2019arXiv190606345L}  \end{scriptsize}\\
\begin{scriptsize} Xinglong-Schmidt           \end{scriptsize} & 
\begin{scriptsize}clear\end{scriptsize}    & 
\begin{scriptsize} $18$    \end{scriptsize}     & 
\begin{scriptsize}$4.5$     \end{scriptsize}  & 
\begin{scriptsize}$0.9$   \end{scriptsize}&
\begin{scriptsize} bayestar ini \end{scriptsize} & 
\begin{scriptsize}$3$ \end{scriptsize} &
\begin{scriptsize} \citet{gcn24190}  \end{scriptsize}\\
\begin{scriptsize} Zwicky  Transient  Facility \end{scriptsize}& 
\begin{scriptsize}g/r-band\end{scriptsize} & 
\begin{scriptsize} $\approx 21$ \end{scriptsize} &
\begin{scriptsize} $1.0$     \end{scriptsize}   &
\begin{scriptsize} $27.8$  \end{scriptsize} & 
\begin{scriptsize}LALInference  \end{scriptsize}&
\begin{scriptsize} $21$  \end{scriptsize} &
\begin{scriptsize} \citet{gcn24191} \end{scriptsize} \\
\hline
\multicolumn{8}{c}{\textbf{S190510g}} \\
\begin{scriptsize} ATLAS     \end{scriptsize} & 
\begin{scriptsize}o-band  \end{scriptsize}   & 
\begin{scriptsize} $19.5$    \end{scriptsize}        & 
\begin{scriptsize}$4.2$  \end{scriptsize}   & 
\begin{scriptsize}$<12$    \end{scriptsize}   & 
\begin{scriptsize}LALinference  \end{scriptsize}& 
\begin{scriptsize}4 \end{scriptsize} &
\begin{scriptsize} \citet{gcn24517} \end{scriptsize} \\
\begin{scriptsize} CNEOST    \end{scriptsize}  & 
\begin{scriptsize}clear \end{scriptsize}     & 
\begin{scriptsize} $\approx 18.5$ \end{scriptsize}    &  
\begin{scriptsize} $9.9$  \end{scriptsize}  & 
\begin{scriptsize}$3.3$   \end{scriptsize} & 
\begin{scriptsize}bayestar ini \end{scriptsize} & 
\begin{scriptsize}13  \end{scriptsize}& 
\begin{scriptsize}  \citet{gcn24465} \end{scriptsize} \\
\begin{scriptsize} Dabancheng/HMT  \end{scriptsize}            & 
\begin{scriptsize}clear  \end{scriptsize}    & 
\begin{scriptsize} $\approx 18$   \end{scriptsize}  & 
\begin{scriptsize} $13.0$   \end{scriptsize} &
\begin{scriptsize} $6.0$  \end{scriptsize}   &
\begin{scriptsize} bayestar ini \end{scriptsize} &
\begin{scriptsize} $\approx 8$ \end{scriptsize} & 
\begin{scriptsize}\citet{gcn24476} \end{scriptsize} \\
\begin{scriptsize} GRAWITA-VST     \end{scriptsize}            & 
\begin{scriptsize}r-sloan \end{scriptsize}   & 
\begin{scriptsize} $22$       \end{scriptsize}      & 
\begin{scriptsize} $21.1$ \end{scriptsize}  & 
\begin{scriptsize}$<6$   \end{scriptsize}     & 
\begin{scriptsize}LALInference  \end{scriptsize}&
\begin{scriptsize} $50$  \end{scriptsize} & 
\begin{scriptsize} \citet{gcn24484}  \end{scriptsize}\\
\begin{scriptsize} GROWTH-DECAM    \end{scriptsize}            & 
\begin{scriptsize}g/r/z-band \end{scriptsize}&
\begin{scriptsize}  $21.7/22.3/21.2$  \end{scriptsize}& 
\begin{scriptsize}$3.0$ \end{scriptsize}  & 
\begin{scriptsize}$18.5$  \end{scriptsize} & 
\begin{scriptsize}LALInference \end{scriptsize} & 
\begin{scriptsize}$65$  \end{scriptsize} & 
\begin{scriptsize}\citet{AnGo2019}  \end{scriptsize}\\
\begin{scriptsize} HSC             \end{scriptsize}     & 
\begin{scriptsize}Y-band  \end{scriptsize}   & 
\begin{scriptsize} $22.7$        \end{scriptsize}   & 
\begin{scriptsize} $<6$   \end{scriptsize}      &
\begin{scriptsize} $<12$     \end{scriptsize} & 
\begin{scriptsize}bayestar ini \end{scriptsize} &  
\begin{scriptsize}  $12$  \end{scriptsize}  &
\begin{scriptsize} \citet{gcn24450} \end{scriptsize} \\
\begin{scriptsize} KMTNet          \end{scriptsize}       & 
\begin{scriptsize}R-band   \end{scriptsize}  &
\begin{scriptsize}  $21.7$      \end{scriptsize}     &
\begin{scriptsize} $13.6$  \end{scriptsize} &
\begin{scriptsize} $<12$  \end{scriptsize}& 
\begin{scriptsize}LALInference  \end{scriptsize}& 
\begin{scriptsize}$66$  \end{scriptsize} &
\begin{scriptsize} \citet{gcn24466}  \end{scriptsize}\\
\begin{scriptsize} MASTER-network  \end{scriptsize}            & 
\begin{scriptsize}clear\end{scriptsize}      & 
\begin{scriptsize} $\approx 18.5$   \end{scriptsize}  & 
\begin{scriptsize} $\approx 0$    \end{scriptsize} & 
\begin{scriptsize} 144  \end{scriptsize}     &
\begin{scriptsize} bayestar ini \end{scriptsize} &
\begin{scriptsize} $52$  \end{scriptsize} &
\begin{scriptsize} \citet{gcn24436} \end{scriptsize} \\
\begin{scriptsize} Pan-STARRS      \end{scriptsize}       & 
\begin{scriptsize}w/i-band    \end{scriptsize}     &
\begin{scriptsize}  $20.5$      \end{scriptsize}     &
\begin{scriptsize} $4.2$   \end{scriptsize}  &
\begin{scriptsize} $<12$  \end{scriptsize}     &
\begin{scriptsize}LALInference  \end{scriptsize}& 
\begin{scriptsize}4  \end{scriptsize}  & 
\begin{scriptsize}  \citet{gcn24517}  \end{scriptsize}\\
\begin{scriptsize} Xinglong-Schmidt\end{scriptsize}            & \begin{scriptsize}clear  \end{scriptsize}    & \begin{scriptsize} $18.5$    \end{scriptsize}       & \begin{scriptsize} $9.8$  \end{scriptsize}  & \begin{scriptsize} $5.8$   \end{scriptsize} & \begin{scriptsize} bayestar ini  \end{scriptsize}&\begin{scriptsize} $19$  \end{scriptsize}& \begin{scriptsize}\citet{gcn24475} \end{scriptsize} \\
\begin{scriptsize}DECAM-KMTNet      \end{scriptsize}      & \begin{scriptsize}r-R band  \end{scriptsize}    & \begin{scriptsize} $>22$     \end{scriptsize}      &\begin{scriptsize} $3.0$    \end{scriptsize}& \begin{scriptsize}$<24$  \end{scriptsize}  &\begin{scriptsize} LALInference \end{scriptsize} &\begin{scriptsize} $69$ \end{scriptsize} &\begin{scriptsize} - \end{scriptsize} \\
\begin{scriptsize}CNEOST-HMT-MASTER-Xinglong-TAROT   \end{scriptsize}         & \begin{scriptsize}clear \end{scriptsize}  & \begin{scriptsize} $>18$  \end{scriptsize}         & \begin{scriptsize}$1.0$ \end{scriptsize}   & \begin{scriptsize}$<24$   \end{scriptsize} & \begin{scriptsize}LALInference \end{scriptsize} &\begin{scriptsize} $71$  \end{scriptsize}& \begin{scriptsize}-  \end{scriptsize}\\
\hline
\multicolumn{8}{c}{\textbf{S190901ap}}  \\
\begin{scriptsize} GOTO           \end{scriptsize}             & \begin{scriptsize}L-band \end{scriptsize}  & \begin{scriptsize} $20$   \end{scriptsize}    & \begin{scriptsize}$0.1$   \end{scriptsize}   & \begin{scriptsize}$54$  \end{scriptsize} & \begin{scriptsize}bayestar ini \end{scriptsize} & \begin{scriptsize} $28$  \end{scriptsize} &\begin{scriptsize} \citet{gcn25654} \end{scriptsize} \\
\begin{scriptsize} GRANDMA-TAROT   \end{scriptsize}              & \begin{scriptsize}clear  \end{scriptsize}  & \begin{scriptsize} $17.5$   \end{scriptsize}      & \begin{scriptsize} $0.4$   \end{scriptsize}    & \begin{scriptsize} $<58.6$ \end{scriptsize} & \begin{scriptsize} LALInference \end{scriptsize} & \begin{scriptsize} $9$ \end{scriptsize}  &\begin{scriptsize}  \citet{gcn25666}  \end{scriptsize} \\
\begin{scriptsize} MASTER-network  \end{scriptsize}            & \begin{scriptsize}clear  \end{scriptsize}    & \begin{scriptsize} $\approx 18.5$ \end{scriptsize}    & \begin{scriptsize} $5.5$   \end{scriptsize}  & \begin{scriptsize} 168   \end{scriptsize}& \begin{scriptsize} bayestar ini  \end{scriptsize}&\begin{scriptsize} $32$   \end{scriptsize}& \begin{scriptsize} \citet{gcn25609} \end{scriptsize} \\
\begin{scriptsize} SVOM-GWAC       \end{scriptsize}         & \begin{scriptsize}R-band\end{scriptsize}   &    \begin{scriptsize}  $16.3$   \end{scriptsize}  & \begin{scriptsize} 12.0  \end{scriptsize} & \begin{scriptsize} 9   \end{scriptsize} &\begin{scriptsize} bayestar ini  \end{scriptsize}&\begin{scriptsize} $16$   \end{scriptsize} &\begin{scriptsize}  \citet{gcn25648}  \end{scriptsize}\\
\begin{scriptsize} Zwicky  Transient  Facility \end{scriptsize}& \begin{scriptsize}g/r-band\end{scriptsize} & \begin{scriptsize} $20.7/20.7$ \end{scriptsize} & \begin{scriptsize} $3.6$ \end{scriptsize}      &\begin{scriptsize} $\sim 72$ \end{scriptsize}  &\begin{scriptsize} LALInference  \end{scriptsize}&\begin{scriptsize} $73$  \end{scriptsize} & \begin{scriptsize}\citet{gcn25616}  \end{scriptsize}\\
\hline
\multicolumn{8}{c}{\textbf{S190910h}} \\
\begin{scriptsize} GRANDMA-TAROT \end{scriptsize}& \begin{scriptsize}clear \end{scriptsize}&\begin{scriptsize}  $18$ \end{scriptsize}& \begin{scriptsize} $10.5$ \end{scriptsize} & \begin{scriptsize} $<129$  \end{scriptsize}&\begin{scriptsize} LALInference  \end{scriptsize}&\begin{scriptsize} $1$  \end{scriptsize}& \begin{scriptsize} \citet{gcn25780}  \end{scriptsize}\\
\begin{scriptsize} MASTER-network  \end{scriptsize}            & \begin{scriptsize}clear\end{scriptsize}      &\begin{scriptsize}  $\approx 18.5$  \end{scriptsize}   & \begin{scriptsize} $2.6$   \end{scriptsize}  & \begin{scriptsize}144  \end{scriptsize} & \begin{scriptsize} bayestar ini \end{scriptsize} & \begin{scriptsize} $8$  \end{scriptsize} &\begin{scriptsize}  \citet{gcn25712} \end{scriptsize} \\
\begin{scriptsize} Zwicky  Transient  Facility \end{scriptsize}& \begin{scriptsize}g/r-band \end{scriptsize}&\begin{scriptsize} $20.7/20.7$ \end{scriptsize}& \begin{scriptsize} $1.80$   \end{scriptsize}   & \begin{scriptsize} $1.5$ \end{scriptsize}  & \begin{scriptsize} bayestar ini \end{scriptsize} & \begin{scriptsize} 34 \end{scriptsize} &\begin{scriptsize} \citet{gcn25722} \end{scriptsize} \\
\hline
\multicolumn{8}{c}{\textbf{S190930t}} \\
\begin{scriptsize} ATLAS       \end{scriptsize}                & \begin{scriptsize}o-band  \end{scriptsize} &\begin{scriptsize}  $19.5$    \end{scriptsize}   & \begin{scriptsize} $0.0$  \end{scriptsize} &\begin{scriptsize} $144$  \end{scriptsize} & \begin{scriptsize}bayestar ini \end{scriptsize} & \begin{scriptsize}$19$  \end{scriptsize} &\begin{scriptsize} \citet{gcn25922}  \end{scriptsize}\\
\begin{scriptsize} MASTER-network  \end{scriptsize}            & \begin{scriptsize}clear   \end{scriptsize}   & \begin{scriptsize} $\approx 18.5$ \end{scriptsize}    & \begin{scriptsize} $\approx 0$   \end{scriptsize}  &\begin{scriptsize} 72  \end{scriptsize} & \begin{scriptsize}bayestar ini  \end{scriptsize}& \begin{scriptsize}$10$  \end{scriptsize} & \begin{scriptsize}\citet{gcn25712}  \end{scriptsize}\\
\begin{scriptsize} Zwicky  Transient  Facility \end{scriptsize}& \begin{scriptsize} g/r-band\end{scriptsize} & \begin{scriptsize} $20.4/20.4$\end{scriptsize} & \begin{scriptsize} $11.9$  \end{scriptsize}  & \begin{scriptsize} $10.0$  \end{scriptsize} &\begin{scriptsize} bayestar ini  \end{scriptsize}&\begin{scriptsize} $45$ \end{scriptsize} &\begin{scriptsize} \citet{gcn25899}  \end{scriptsize}\\
\hline
\enddata
\end{deluxetable*}

\begin{deluxetable*}{cccccccc}
\tablenum{3}
\setlength{\tabcolsep}{0pt}
\tablecaption{Reports of the observations by various teams of the sky localization area of gravitational-wave alerts of possible BHNS candidates S190426c, S190814bv, S190910d and S190923y. Teams that employed ``galaxy targeting'' during their follow-up are not mentioned here. In the case where numbers were not reported or provided upon request, we recomputed some of them; if this was not possible, we add $-$.}
\label{tab:TableobsNSBH}
\tablewidth{0pt}
\tablehead{\begin{scriptsize}Telescope\end{scriptsize} & \begin{scriptsize}Filter\end{scriptsize} & \begin{scriptsize}Limit mag\end{scriptsize} & \begin{scriptsize}Delay aft. GW \end{scriptsize} & \begin{scriptsize}Duration\end{scriptsize} & \multicolumn{2}{c}{\begin{scriptsize}GW sky localization area\end{scriptsize}}  & \begin{scriptsize}reference\end{scriptsize} \\  &  &  & \begin{scriptsize}(h)\end{scriptsize} & \begin{scriptsize}(h)\end{scriptsize} & \begin{scriptsize}name\end{scriptsize} & \begin{scriptsize}coverage (\%)\end{scriptsize} & }
\startdata
\multicolumn{8}{c}{\textbf{S190426c}} \\
\begin{scriptsize} ASAS-SN         \end{scriptsize}            & \begin{scriptsize}g-band \end{scriptsize}  & \begin{scriptsize}$\approx 18$ \end{scriptsize}  & \begin{scriptsize} - \end{scriptsize}  & \begin{scriptsize}$\approx 24$ \end{scriptsize}& \begin{scriptsize}bayestar ini\end{scriptsize} &\begin{scriptsize} $86$ \end{scriptsize} &\begin{scriptsize} \citet{gcn24309} \end{scriptsize} \\
\begin{scriptsize} CNEOST           \end{scriptsize}           & \begin{scriptsize}clear    \end{scriptsize}& \begin{scriptsize}$\approx 20$ \end{scriptsize}  & \begin{scriptsize}$1.3$ \end{scriptsize} & \begin{scriptsize}$3.5$ \end{scriptsize}      & \begin{scriptsize}bayestar ini\end{scriptsize} &\begin{scriptsize} $35$\end{scriptsize} & \begin{scriptsize} \citet{gcn24286}\end{scriptsize} \\
\begin{scriptsize} DDOTI/OAN        \end{scriptsize}           & \begin{scriptsize}w-band  \end{scriptsize} & \begin{scriptsize}$\approx 18.5$ \end{scriptsize}         & \begin{scriptsize}$14.7$\end{scriptsize} & \begin{scriptsize}$4.9$        \end{scriptsize}& \begin{scriptsize}bayestar ini\end{scriptsize} &\begin{scriptsize} $\approx 37$ \end{scriptsize}&\begin{scriptsize} \citet{gcn24310} \end{scriptsize}\\
\begin{scriptsize} GOTO             \end{scriptsize}           & \begin{scriptsize}g-band \end{scriptsize}  & \begin{scriptsize}$19.9$ \end{scriptsize}        &\begin{scriptsize} $5.3$ \end{scriptsize} & \begin{scriptsize}$8.9$     \end{scriptsize}   & \begin{scriptsize}LALInference \end{scriptsize}& \begin{scriptsize}$54$  \end{scriptsize}& \begin{scriptsize}\citet{gcn24291}\end{scriptsize} \\
\begin{scriptsize} GRANDMA-OAJ      \end{scriptsize}           &\begin{scriptsize} r-band \end{scriptsize}  &\begin{scriptsize} $19.6$  \end{scriptsize}       & \begin{scriptsize}$6.3$ \end{scriptsize}  & \begin{scriptsize}$4.9$        \end{scriptsize}& \begin{scriptsize}bayestar ini\end{scriptsize} & \begin{scriptsize}$11$ \end{scriptsize} & \begin{scriptsize}\citet{gcn24327}\end{scriptsize} \\
\begin{scriptsize} GRAWITA-Asiago   \end{scriptsize}           & \begin{scriptsize}r-band \end{scriptsize}  &\begin{scriptsize} $\approx 16$\end{scriptsize}   &\begin{scriptsize} $6.9$ \end{scriptsize} & \begin{scriptsize}$0.5$     \end{scriptsize}  & \begin{scriptsize}LALInference\end{scriptsize} &\begin{scriptsize} $2$\end{scriptsize}   &\begin{scriptsize} \citet{gcn24340}\end{scriptsize} \\
\begin{scriptsize} GROWTH-DECAM     \end{scriptsize}           &\begin{scriptsize} r/z-band\end{scriptsize} & \begin{scriptsize}$22.9/22.5$\end{scriptsize}    &\begin{scriptsize} $7.6$ \end{scriptsize} & \begin{scriptsize}$11.5$    \end{scriptsize}  & \begin{scriptsize}LALInference \end{scriptsize}& \begin{scriptsize}$8.0$\end{scriptsize} & \begin{scriptsize}\citet{GoAn2019}\end{scriptsize} \\
\begin{scriptsize} GROWTH-Gattini-IR    \end{scriptsize}       & \begin{scriptsize}J-band  \end{scriptsize} &\begin{scriptsize} $\approx 14.5$\end{scriptsize} &\begin{scriptsize} $11.8$ \end{scriptsize}& \begin{scriptsize}$9.8$     \end{scriptsize}  & \begin{scriptsize}bayestar ini\end{scriptsize} &\begin{scriptsize} $92$  \end{scriptsize}&\begin{scriptsize} \citet{gcn24284}\end{scriptsize} \\
\begin{scriptsize} GROWTH-INDIA        \end{scriptsize}        & \begin{scriptsize}r-band \end{scriptsize}  &\begin{scriptsize} $20.5$ \end{scriptsize}        & \begin{scriptsize}$2.0$ \end{scriptsize}  & \begin{scriptsize}$29.4$ \end{scriptsize}   &\begin{scriptsize} bayestar ini\end{scriptsize} &\begin{scriptsize} $4$ \end{scriptsize}  &\begin{scriptsize} \citet{gcn24258} \end{scriptsize}\\
\begin{scriptsize} J-GEM              \end{scriptsize}         &\begin{scriptsize} clear\end{scriptsize}    & \begin{scriptsize}$20$\end{scriptsize}           & \begin{scriptsize}$19$ \end{scriptsize}   & \begin{scriptsize}-         \end{scriptsize}  & \begin{scriptsize}bayestar ini     \end{scriptsize}    & \begin{scriptsize}- \end{scriptsize}   & \begin{scriptsize}\citet{gcn24299}\end{scriptsize} \\
\begin{scriptsize} MASTER-network     \end{scriptsize}         & \begin{scriptsize}clear    \end{scriptsize}& \begin{scriptsize}$\approx 18.5$\end{scriptsize}   &\begin{scriptsize} $\approx 0$\end{scriptsize}   &\begin{scriptsize} $144$    \end{scriptsize}    & \begin{scriptsize}bayestar\end{scriptsize} ini &\begin{scriptsize} $53$ \end{scriptsize} &\begin{scriptsize} \citet{gcn24236}\end{scriptsize} \\
\begin{scriptsize} SAGUARO            \end{scriptsize}         & \begin{scriptsize}g-band  \end{scriptsize} &\begin{scriptsize} $\approx 21$\end{scriptsize}   &\begin{scriptsize} 41.8  \end{scriptsize}  &\begin{scriptsize} $\approx 24$\end{scriptsize} & \begin{scriptsize}bayestar ini \end{scriptsize}& \begin{scriptsize}$5$  \end{scriptsize} &\begin{scriptsize} \citet{2019arXiv190606345L} \end{scriptsize}\\
\begin{scriptsize} Zwicky  Transient  Facility\end{scriptsize} &\begin{scriptsize} g/r-band\end{scriptsize} &\begin{scriptsize} $\approx 21$ \end{scriptsize}  &\begin{scriptsize} $13.0$ \end{scriptsize} & \begin{scriptsize}$31.3$  \end{scriptsize}    &\begin{scriptsize} bayestar ini\end{scriptsize} &\begin{scriptsize} $75$ \end{scriptsize} &\begin{scriptsize} \citet{gcn24283}\end{scriptsize} \\
\hline
\multicolumn{8}{c}{\textbf{S190814bv}} \\
\begin{scriptsize} ATLAS              \end{scriptsize}         & \begin{scriptsize}o-band \end{scriptsize}    & \begin{scriptsize}$>16$     \end{scriptsize}      & \begin{scriptsize}$<12$ \end{scriptsize}   &\begin{scriptsize} $24.0$  \end{scriptsize}     & \begin{scriptsize}LALinference \end{scriptsize}& \begin{scriptsize}$99$ \end{scriptsize}  & \begin{scriptsize}\citet{gcn25375}\end{scriptsize} \\
\begin{scriptsize} DESGW-DECam        \end{scriptsize}        & \begin{scriptsize}r/i-band\end{scriptsize} & \begin{scriptsize}$23.4,22.6$ \end{scriptsize}   & \begin{scriptsize}$9.5$ \end{scriptsize} & \begin{scriptsize}$\approx 96$    \end{scriptsize}  & \begin{scriptsize}LALInference \end{scriptsize}& \begin{scriptsize}$90$ \end{scriptsize}& \begin{scriptsize}\citet{gcn25336}\end{scriptsize} \\
\begin{scriptsize} DDOTI/OAN          \end{scriptsize}         & \begin{scriptsize}w-band \end{scriptsize}  & \begin{scriptsize}$\approx 18.5$  \end{scriptsize}        & \begin{scriptsize}$10.8$\end{scriptsize} &\begin{scriptsize} $3.9$        \end{scriptsize}& \begin{scriptsize}LALInference\end{scriptsize} & \begin{scriptsize}$ 90$\end{scriptsize} &  \begin{scriptsize}\citet{gcn25352}\end{scriptsize} \\
\begin{scriptsize} GOTO               \end{scriptsize}         &\begin{scriptsize} L-band \end{scriptsize}  & \begin{scriptsize}$18.4$  \end{scriptsize}     & \begin{scriptsize}$3.5$   \end{scriptsize}  &\begin{scriptsize} $5.0$ \end{scriptsize} &\begin{scriptsize} LALInference\end{scriptsize} &\begin{scriptsize} $83$ \end{scriptsize} & \begin{scriptsize}\citet{gcn25337} \end{scriptsize}\\
\begin{scriptsize} GRAWITA-VST        \end{scriptsize}         & \begin{scriptsize}r-sloan  \end{scriptsize}  & \begin{scriptsize}$\approx 22$ \end{scriptsize}            & \begin{scriptsize}$11.7$ \end{scriptsize} &\begin{scriptsize} $1.3$     \end{scriptsize}  &\begin{scriptsize} LALInference \end{scriptsize}&\begin{scriptsize} $65$\end{scriptsize}  & \begin{scriptsize}\citet{gcn25371} \end{scriptsize}\\
\begin{scriptsize} GROWTH-Gattini-IR  \end{scriptsize}         & \begin{scriptsize}J-band \end{scriptsize}  &\begin{scriptsize} $\approx 17.0$\end{scriptsize} &\begin{scriptsize} $8.9$\end{scriptsize} & \begin{scriptsize}$96.0$     \end{scriptsize}  & \begin{scriptsize}bayestar ini\end{scriptsize} &\begin{scriptsize} $90$\end{scriptsize}  &\begin{scriptsize} \citet{gcn25358}\end{scriptsize} \\
\begin{scriptsize} KMTNet             \end{scriptsize}         & \begin{scriptsize}R-band \end{scriptsize}    &\begin{scriptsize} $22.0$    \end{scriptsize} & \begin{scriptsize}$8.4$\end{scriptsize}  & \begin{scriptsize}$<12$\end{scriptsize}  & \begin{scriptsize}LALinference\end{scriptsize} &\begin{scriptsize} $98$\end{scriptsize} & \begin{scriptsize}\citet{gcn25342} \end{scriptsize}\\
\begin{scriptsize} MASTER-network     \end{scriptsize}         & \begin{scriptsize}clear    \end{scriptsize}  & \begin{scriptsize}$\approx 18$\end{scriptsize}     & \begin{scriptsize}$0.4$ \end{scriptsize}   &\begin{scriptsize} $6.7$ \end{scriptsize}     & \begin{scriptsize}bayestar ini \end{scriptsize}&\begin{scriptsize} $98$\end{scriptsize}  &\begin{scriptsize} \citet{gcn25322} \end{scriptsize}\\
\begin{scriptsize} MeerLICHT          \end{scriptsize}         &\begin{scriptsize} u/q/i-band \end{scriptsize}  &\begin{scriptsize} $18.5/19.7/19.1$\end{scriptsize}       &\begin{scriptsize} $2.0$\end{scriptsize}    & \begin{scriptsize}$5.1$ \end{scriptsize}  & \begin{scriptsize}bayestar HLV\end{scriptsize} &\begin{scriptsize} $95$ \end{scriptsize} & \begin{scriptsize}\citet{gcn25340} \end{scriptsize}\\
\begin{scriptsize} Pan-STARRS         \end{scriptsize}    & \begin{scriptsize}i/z-band  \end{scriptsize}       & \begin{scriptsize}$20.8/20.3$\end{scriptsize}           & \begin{scriptsize}$15.5$\end{scriptsize}    &\begin{scriptsize} $2.55$  \end{scriptsize}     & \begin{scriptsize}LALInference\end{scriptsize} &\begin{scriptsize} 89\end{scriptsize} &\begin{scriptsize} \citet{gcn24517}\end{scriptsize} \\
\begin{scriptsize} Swope              \end{scriptsize}         &\begin{scriptsize} r-band \end{scriptsize}& \begin{scriptsize}$\approx 20.0$ \end{scriptsize}  &\begin{scriptsize} $6.3$\end{scriptsize}  &\begin{scriptsize} $5.1$ \end{scriptsize}     &\begin{scriptsize} LALinference\end{scriptsize} &\begin{scriptsize} $42$ \end{scriptsize} &\begin{scriptsize} \citet{gcn25350}\end{scriptsize} \\
\begin{scriptsize} GRANDMA-TCA       \end{scriptsize}      & \begin{scriptsize}clear \end{scriptsize}   &\begin{scriptsize} $18.0$     \end{scriptsize}  & \begin{scriptsize}$3.0$ \end{scriptsize}   & \begin{scriptsize}$2.5$  \end{scriptsize} &\begin{scriptsize} bayestar HLV\end{scriptsize} &\begin{scriptsize} $27$\end{scriptsize}  &\begin{scriptsize} \citet{gcn25338} \end{scriptsize}\\
\begin{scriptsize} GRANDMA-TRE        \end{scriptsize}      &\begin{scriptsize} clear \end{scriptsize}   &\begin{scriptsize} $17.0$ \end{scriptsize}      & \begin{scriptsize}$0.5$  \end{scriptsize}  & \begin{scriptsize}$1.0$ \end{scriptsize}  &\begin{scriptsize} LALinference\end{scriptsize} &\begin{scriptsize} $76$ \end{scriptsize} &\begin{scriptsize} \citet{gcn25599} \end{scriptsize}\\
\begin{scriptsize} Zwicky Transient Facility \end{scriptsize}  &\begin{scriptsize} g/r/i-band \end{scriptsize}& \begin{scriptsize}$20.3$\end{scriptsize}   &\begin{scriptsize} $13.3$  \end{scriptsize}&\begin{scriptsize} $1.7$\end{scriptsize}      & \begin{scriptsize}bayestar ini HLV \end{scriptsize}&\begin{scriptsize} $86$\end{scriptsize}  &\begin{scriptsize} \citet{gcn25343} \end{scriptsize}\\
\begin{scriptsize} MASTER-TAROT  \end{scriptsize} &\begin{scriptsize} clear \end{scriptsize}& \begin{scriptsize}$>17$   \end{scriptsize}&\begin{scriptsize} $0.5$\end{scriptsize}  & \begin{scriptsize}$<3$ \end{scriptsize}     & \begin{scriptsize}LALInference\end{scriptsize} & \begin{scriptsize}$89$ \end{scriptsize} & \begin{scriptsize}-\end{scriptsize} \\
\hline
\multicolumn{8}{c}{\textbf{S190910d}} \\
\begin{scriptsize} DDOTI/OAN        \end{scriptsize}   & \begin{scriptsize}w-band \end{scriptsize}   & \begin{scriptsize}$\approx 18.5$ \end{scriptsize}  & \begin{scriptsize}$3.0$\end{scriptsize}   &\begin{scriptsize} $4.7$     \end{scriptsize}   & \begin{scriptsize}LALInference \end{scriptsize}& \begin{scriptsize}$5$ \end{scriptsize} &\begin{scriptsize} \citet{gcn25737}\end{scriptsize} \\
\begin{scriptsize} GRANDMA-TAROT      \end{scriptsize}     & \begin{scriptsize}clear    \end{scriptsize}& \begin{scriptsize}$\approx 17.5$ \end{scriptsize}  &\begin{scriptsize} $1.0$\end{scriptsize}   &\begin{scriptsize} $<67$ \end{scriptsize}       & \begin{scriptsize}LALInference\end{scriptsize} &\begin{scriptsize} $37$\end{scriptsize}  &\begin{scriptsize} \citet{gcn25749}\end{scriptsize} \\
\begin{scriptsize} MASTER-network      \end{scriptsize}        &\begin{scriptsize} clear    \end{scriptsize}& \begin{scriptsize} $\approx 18.5$ \end{scriptsize}  &\begin{scriptsize} $\approx 0$ \end{scriptsize}  &\begin{scriptsize} $144$    \end{scriptsize}    & \begin{scriptsize}bayestar ini \end{scriptsize}&\begin{scriptsize} $25$ \end{scriptsize} & \begin{scriptsize}\citet{gcn25694} \end{scriptsize}\\
\begin{scriptsize} Zwicky Transient Facility \end{scriptsize}  & \begin{scriptsize}g/r-band \end{scriptsize}&\begin{scriptsize} $20.8$ \end{scriptsize}  &\begin{scriptsize} $1.5$\end{scriptsize}  &\begin{scriptsize}  $1.5$\end{scriptsize}   & \begin{scriptsize} bayestar ini\end{scriptsize} &\begin{scriptsize} $34$ \end{scriptsize} &\begin{scriptsize} \citet{gcn25706}\end{scriptsize} \\
\hline
\multicolumn{8}{c}{\textbf{S190923y}}  \\
\begin{scriptsize} GRANDMA-TAROT      \end{scriptsize}     & \begin{scriptsize}clear\end{scriptsize}    & \begin{scriptsize}$\approx 17.5$\end{scriptsize}   & \begin{scriptsize}$3.7$ \end{scriptsize}  & \begin{scriptsize}$<56.1$    \end{scriptsize}    &\begin{scriptsize} bayestar ini \end{scriptsize}&\begin{scriptsize} $26$ \end{scriptsize} &\begin{scriptsize} \citet{gcn25847}\end{scriptsize} \\
\begin{scriptsize} MASTER-network     \end{scriptsize}         & \begin{scriptsize}clear    \end{scriptsize}& \begin{scriptsize}$\approx 18.5$ \end{scriptsize}  &\begin{scriptsize} $\approx 0$  \end{scriptsize} &\begin{scriptsize} $144$    \end{scriptsize}    &\begin{scriptsize} bayestar ini \end{scriptsize}&\begin{scriptsize} $58$ \end{scriptsize} &\begin{scriptsize} \citet{gcn25812}\end{scriptsize} \\
\hline
\enddata
\end{deluxetable*}

\end{document}